\documentclass[aps,prb,amsmath,twocolumn,amssymb,titlepage,superscriptaddress,showpacs]{revtex4-2}
\usepackage[english]{babel}
\usepackage{graphicx} 
\usepackage{transparent}
\usepackage{amsmath}
\usepackage{amssymb}
\usepackage{epstopdf}
\usepackage{amsfonts}
\usepackage{bm}
\usepackage{times}
\usepackage{color}
\usepackage{mathtools}
\usepackage{hyperref}
\usepackage{tabularx}
\usepackage{hhline}

\newcommand{\ket}[1]{|#1\rangle}

\newcommand{\matele}[3]{\langle #1|#2|#3\rangle}
\newcommand{\I}{\mathrm{i}}

\newcommand{\ave}[1]{\langle #1 \rangle}

\newcommand{\astcycl}{\mathrlap{\kern0.085em{\circlearrowright}}\ast}
\newcommand{\taucycl}{\mathrlap{\kern0.42em{\bullet}}\circlearrowright}

\begin{document}

\title{Measuring the ultrafast screening of $U$ in photo-excited charge-transfer insulators with time-resolved  X-ray absorption spectroscopy }
\author{Denis Gole\v z}
\affiliation{Jozef Stefan Institute, Jamova 39, SI-1000 Ljubljana, Slovenia}
\affiliation{Faculty of Mathematics and Physics, University of Ljubljana, Jadranska 19, 1000 Ljubljana, Slovenia}
\author{Eva Paprotzki}
\affiliation{I. Institute for Theoretical Physics, University of Hamburg, Notkestraße 9-11, 22607 Hamburg, Germany}
\affiliation{The Hamburg Centre for Ultrafast Imaging, Hamburg, Germany}
\author{Philipp Werner}
\affiliation{Department of Physics, University of Fribourg, 1700 Fribourg, Switzerland}
\author{Martin Eckstein}
\affiliation{I. Institute for Theoretical Physics, University of Hamburg, Notkestraße 9-11, 22607 Hamburg, Germany}
\affiliation{The Hamburg Centre for Ultrafast Imaging, Hamburg, Germany}

\begin{abstract}
Recent seminal experiments have utilized time-resolved X-ray absorption spectroscopy (XAS) to investigate the ultrafast photo-induced renormalization of the electron interaction (``Hubbard $U$'') in Mott and charge transfer insulators.
In this paper, we analyze the change of interactions due to dynamical screening as it is encoded in the XAS signal, using the non-equilibrium GW+EDMFT formalism. Our study shows that XAS is well-suited for measuring this change,
but two aspects should be kept in mind if the screening processes are not substantially faster than the valence electron dynamics:  (i) Screening in a photo-excited system can affect both the position and the lineshape of the absorption lines. (ii) In general, the effect cannot be captured by the modification of a single interaction parameter. Specifically, an estimate for $\Delta U$ extracted from the shift of the XAS lines does not necessarily describe the related shift of the the upper Hubbard band. We clarify these aspects using a minimal cluster model and the three-band Emery model for a charge transfer insulator.
\end{abstract}

\maketitle

\section{Introduction}

Due to its element selectivity, X-ray absorption spectroscopy (XAS) is a highly suitable probe of the local electronic structure in complex materials~\cite{de20212p,ament2011,kuo2017challenges}. State-of-the-art X-ray free electron laser sources provide sufficient energy and time resolution to measure local multiplet energies on femtosecond timescales, which is useful to guide the engineering of functional material  properties under non-equilibrium conditions~\cite{Giannetti2016,Basov2017,torre2021,Murakami2023,boschini2024}. In particular, XAS can potentially  unravel photo-induced level shifts and elucidate the orbital nature of mobile carriers after ``photo-doping'', i.e., a laser-induced transfer of electrons between different bands \cite{baykusheva2022,wang2022ultrafast,graanas2022}. Photo-doping in  Mott- or charge transfer insulators has been predicted to induce non-equilibrium phases ranging from hidden magnetic and orbital orders to superconductivity \cite{Murakami2023}. As is well-known from semiconductor studies \cite{huber2001}, photo-doping can also modify the dielectric properties of a material, and therefore lead to a rapid screening of the electron interaction. In strongly correlated electron systems, this provides an intriguing possibility to control effective interaction parameters such as the local Hubbard interaction $U$ \cite{golez2019,golez2019a,tancogne2018}, with potentially profound impacts on the many-body state. Recently recorded shifts in the X-ray absorption spectrum of a photo-excited cuprate and nickelates  can indeed be interpreted in terms of an ultrafast modification of $U$ \cite{baykusheva2022,graanas2022,wang2022ultrafast}, and similar shifts have been observed in photo-doped NiO~\cite{lojewski2024}. 

The analysis of time-resolved XAS needs a robust theoretical modeling, because it relies on many relevant parameters, such as band shifts due to the interaction of electrons with photo-doped holes, and the core valence interaction. This has motivated  microscopic theoretical descriptions of time-resolved XAS based on exact diagonalization~\cite{chen2019,baykusheva2022} and nonequilibrium  Green's functions~\cite{werner2022}.  In the present work, we will address a particular conceptual difficulty: Screening is not instantaneous. The photo-induced renormalization of the interaction originates from charge fluctuations which are activated by photo-doping, and which have energies comparable to the energy scales in the valence band. One therefore needs to clarify whether and how the dynamics of the screening process affects the XAS signal. 

A theoretical description of screening in 
correlated electron systems 
can be achieved by combining the GW formalism \cite{hedin1965} with extended dynamical mean-field theory (EDMFT) \cite{sun2002,ayral2013,golez2015,nilsson2017,boehnke2016}. In this approach, similar to DMFT \cite{georges1996}, the system is mapped to an impurity model which embeds one lattice site in a self-consistent environment that describes the rest of the lattice. In GW+EDMFT, this environment includes a fermion reservoir, with which the impurity can exchange electrons,  and a spectrum of bosonic modes, which represent charge fluctuations in the lattice that can screen the local interaction.  DMFT is well suited for calculating XAS spectra, because the core level can be added to the effective impurity model without changing the self-consistent environment \cite{cornaglia2007, haverkort2014, luder2017, hariki2018}, also in the time-dependent case~\cite{werner2022}.  In the present work, we extend this  idea to GW+EDMFT, in order to analyze how dynamic and static screening processes are reflected in time-resolved XAS. 

The paper is structured as follows: In Sec.~\ref{Sec:XASembedding} we explain the evaluation of XAS within time-dependent GW+EDMFT simulations. In Sec.~\ref{Sec:XAS_har}, we demonstrate the effect of dynamical screening using a minimal model with a single bosonic mode to represent the charge fluctuations. Finally, in Sec.~\ref{sec:results} we discuss the photo-induced changes of the XAS spectrum within a model for a photo-doped charge transfer insulator. Sec.~\ref{Sec:conclusion} provides a summary and further discussions.

\section{Evaluation of XAS within GW+EDMFT}
\label{Sec:XASembedding}

\subsection{General expression for the XAS signal}
\label{ssec:XASimp}

To calculate the X-ray absorption spectrum within the GW+EDMFT formalism, we closely follow the discussion in Ref.~\onlinecite{werner2022}. We will first summarize the main aspects of the general formalism (without repeating the derivation), before discussing details relevant for GW+EDMFT in Sec.~\ref{GWEDMFT} and \ref{XASGWEDMFT}.  

We start from a generic lattice model, with a set of valence and core orbitals at each lattice site $j$. The creation/annihilation operators for the valence and core electrons will be denoted by $d_{j,\alpha}$ ($d_{j,\alpha}^\dagger$) and $c_{j,\alpha}$ ($c_{j,\alpha}^\dagger$), respectively; Greek indices $\alpha$ label orbital and spin.  For simplicity, the core-level Hamiltonian represents a single level with energy $\epsilon_c<0$, 
\begin{align}
H_{c}=\epsilon_c \sum_{j,\alpha} c_{j,\alpha}^\dagger c_{j,\alpha},
\label{hc01}
\end{align} 
although more structure could be added~\cite{tanaka1994,haverkort2012,sipr2011,luder2017}. Within a semiclassical treatment of the incoming  X-ray photon beam, a dipolar excitation from the core to the valence $d$-orbitals is described by the Hamiltonian 
\begin{align}
H_{\text{dip}}=g\big(s(t)e^{-\I\omega_{\text{in}} t}+h.c. \big) \sum_{j} \big(P_{j}+P_{j}^\dagger\big),
\label{hdip01}
\end{align} 
where $s(t)$ is the probe envelope, $\omega_{\text{in}}$ is the frequency of the incoming X-ray pulse, and $P_{j}^\dagger = \sum_{\alpha,\alpha'} D_{\alpha,\alpha'} \,d^\dagger_{j,\alpha}c_{j,\alpha'} $ the dipolar transition operator, with the matrix element  $D_{\alpha,\alpha'}=\matele{d,\alpha}{\vec{r}\cdot\vec{\epsilon}}{c,\alpha'}$ of the displacement operator ($\vec{r}$) projected onto the X-ray polarization direction $\vec{\epsilon}$. Due to the strong  localization of the core orbital, we assume that the dipolar transition is between orbitals at the same site $j$ only.  The prefactor $g$ sets the overall amplitude of the incoming probe pulse. 

Because the X-ray photon energy  $\omega_{\text{in}}$ is much larger than the energy of the valence electrons,
we employ the rotating wave approximation and replace the dipolar transition \eqref{hdip01} by 
\begin{align}
H_{\text{dip}}=  g
\sum_{j}
\Big(s(t)e^{-\I \omega_{\text{in}} t} P_{j}^\dagger + h.c.
\Big),
\label{hdip02}
\end{align}
where terms $\sim e^{-\I\omega _{\text{in}}t} P_j$, which combine the emission of a photon and the creation of a core hole, are neglected. The X-ray signal is the fraction of absorbed photons during the pulse, or, in the semiclassical description, the ratio of the absorbed energy and the photon energy $\omega_{\text{in}}$. It is given by \cite{werner2022}
\begin{align}
	I_{\text{XAS}}=
	\lim_{g\to 0} \frac{1}{g^2}
	\sum_{j}
	\int
	dt \Big[\,gs(t) e^{-\I \omega_{\text{in}}t}\ave{P_{j}^\dagger(t)} - h.c.\Big].
	\label{eq:I_XAS_IMP1}
\end{align}
The limit $g\to0$ indicates that the absorption is to be calculated to leading order in $g$, where the time-dependent expectation value 
$\ave{P_{j}^\dagger(t)}$ evaluated in the presence of the probe pulse~\eqref{hdip02} is of the order $\mathcal{O}(g)$. In the simulations, we explicitly evaluate the absorption signal $I_{\text{XAS}}$ with a nonzero probe pulse, and take the limit  $g\to 0$ numerically.

An important observation is that Eq.~\eqref{eq:I_XAS_IMP1} is a sum of local terms. One can therefore formally obtain the XAS signal by evaluating Eq.~\eqref{eq:I_XAS_IMP1} with an effective action 
 \begin{align}
 S_{j}= S_{\text{loc}} + S_{\text{env}}
 \label{actionfuxes2}
 \end{align}
 for site $j$, where $S_{\text{loc}}$ describes the dynamics generated by the local Hamiltonian $H_{\text{loc}}$ for the core and valence orbitals on site $j$, while $S_{\text{env}}$ describes the effect of the environment and is formally obtained by integrating out all degrees of freedom of the remaining lattice. Embedding techniques based on DMFT map the lattice model to a quantum impurity model which can reproduce the local correlation functions at a given site $j$, and thus yield an explicit expression for $S_{\text{env}}$.
Because the XAS signal itself is already of order $\mathcal{O}(g^2)$ in the probe amplitude $g$, one can to leading order in $g$ first determine the environment action $S_{\text{env}}$ in a simulation without the core levels, and 
then 
compute XAS in a post-processing step from an impurity model that has the same $S_{\text{env}}$, but is extended by the core level and the local part of the dipolar interaction~\eqref{hdip02}. 

\subsection{GW+EDMFT}
\label{GWEDMFT}

The valence electrons interact with a general density-density interaction,
\begin{align}
\label{Hintgeneral}
H_{\text{int}} = \frac{1}{2}\sum_{j,j'}\sum_{\alpha,\alpha'} 
n_{d,j \alpha} v_{j-j'}^{\alpha,\alpha'} n_{d,j'\alpha'}. 
\end{align}
Here $n_{d,j\alpha}=d_{j\alpha}^\dagger d_{j\alpha}$ is the local spin/orbital occupation number, $v_{j-j'}$ denotes the bare interaction in real space, and $v_{\bm q}$ its Fourier transform (all interactions are understood as matrices in spin/orbital indices).  The effect of screening is encoded in the fully screened interaction $W_{\bm q}(t,t')$. The latter is the bare interaction reduced by the screening function (inverse dielectric function), 
\begin{align}
\label{Wq1}
W_{\bm q} =  \epsilon_{\bm q}^{-1} \ast v_{\bm q},
\end{align}
which in turn satisfies the exact relation 
\begin{align}
\label{Wq2}
\epsilon_{\bm q}^{-1} = 1-v_{\bm q} \ast \chi_{\bm q},
\end{align}
with the charge correlation function defined as $\chi_{\bm q}(t,t') = -i \langle T_\mathcal{C} n_{d,\bm q}(t)n_{d,-\bm q}(t) \rangle$.
Here and below, the functions are understood as matrices in spin/orbital indices, and for correlation functions, we use the two-time notation for Keldysh Green's functions, following Ref.~\onlinecite{aoki2014_rev}; $T_\mathcal{C}$ is the time ordering operator on the Keldysh contour. 
In equilibrium or in a steady state, the arguments $(t,t')$ can be replaced by a single frequency. Products $A\ast B$ correspond to a convolution in time (product in frequency) and a matrix multiplication in the orbital indices.

Within the GW+EDMFT formalism, both the local Green's function $G_{\text{loc}}=\frac{1}{N_{\bm k}}\sum_{\bm k} G_{\bm k}$  and the local fully screened interaction $W_{\text{loc}}=\frac{1}{N_{\bm k}}\sum_{\bm k} W_{\bm k}$ are obtained  from a quantum impurity model. (All sites are equivalent due to translational invariance.) The impurity action  is of the form of Eq.~\eqref{actionfuxes2}, with 
\begin{align}
\label{SGWD}
S_{\text{env}}=S_\Delta+S_{\text{mf}} + S_{\mathcal{D}}.
\end{align}
Here  $S_{\Delta}=-\sum_{\alpha\alpha'}\int dtdt'\,d_\alpha^*(t) \Delta_{\alpha,\alpha'}(t,t') d_{\alpha'}(t') \equiv - d^\dagger \ast \Delta \ast d$ describes the tunnelling of the valence electrons in and out of the lattice site $j$ via the hybridization function $\Delta$, and $S_{\text{mf}} $ is a mean-field potential for the 
electrons on site $j$ due to the static interaction with other sites (Hartree self energy). The last term is a retarded interaction
\begin{align}
& S_\mathcal{D}=- \frac{1}{2} \sum_{j,j'} \sum_{\alpha,\alpha'} \, n_{d,j\alpha}\ast \mathcal{D} \ast n_{d,\alpha'j'},
\label{envU}
\end{align}
which represents a correction to the bare and instantaneous on-site interaction; we will denote the full dynamical impurity interaction by $\mathcal{U}=U+\mathcal{D}$. The functions $\Delta$ and $\mathcal{D}$ are determined through self-consistency relations, whose precise form is given in Ref.~\onlinecite{golez2019,golez2019a}. In particular, self-consistency enforces that the local fully screened interaction $W_{\text{loc}}$ matches the fully screened interaction  $W_{\text{imp}}$ of the impurity model, which satisfies an equation $W_{\text{imp}} = \mathcal{U} - \mathcal{U}\ast \chi_{\text{imp}} \ast \mathcal{U}$ that is a local analog to Eqs.~\eqref{Wq1} and \eqref{Wq2}, with $\chi_{\text{imp}}$ the local charge susceptibility of the impurity model.

Within GW+EDMFT, the effect of screening can therefore be seen both in the fully screened interaction $W_{\text{imp}}=W_{\text{loc}}$, and in the effective impurity interaction $\mathcal{U}$. It should be emphasized, though, that neither quantity has the meaning of an effective Hubbard interaction in the lattice model. In particular, $\mathcal{U}$ is an auxiliary quantity used to characterize local correlation functions, and an impurity model with an effective interaction $\mathcal{U}$ does not imply that the system behaves like a Hubbard model with interaction $\mathcal{U}$ at every site. 

\subsection{XAS from GW+EDMFT}
\label{XASGWEDMFT}

To obtain the XAS signal within the GW+EDMFT formalism, one first performs a time-dependent simulation without the core level. This fixes $\Delta$ and $\mathcal{U}$ of the impurity action. In a second step, one adds the core level and the dipolar excitation \eqref{hdip02} to the local Hamiltonian, and computes the XAS signal from Eq.~\eqref{eq:I_XAS_IMP1}. The numerical computation of the XAS signal within an impurity model with retarded interaction still presents a formidable challenge. We use the strong coupling expansion approach of Ref.~\onlinecite{golez2015}, which implements a double expansion in $\mathcal{D}$ and $\Delta$. For non-equilibrium steady-states, an exact evaluation of the strong-coupling expansion has recently been developed \cite{erpenbeck2023} and used in the context of non-equilibrium DMFT \cite{kunzel2024}, but so far without retarded interactions. The development of a corresponding exact impurity solver for retarded interactions is beyond the scope of this work, and we restrict ourselves to the leading non-crossing approximation (NCA), as in previous GW+EDMFT studies \cite{golez2019,golez2019a}. In part, the accuracy of this expansion will be assessed by comparison to the exactly solvable model of Sec.~\ref{Sec:XAS_har}.

The  core hole has typically a short lifetime of a few femtoseconds, due to decay processes originating from  Auger-Meitner decay~\cite{de20212p}. Following previous theoretical works~\cite{cornaglia2007, haverkort2014, luder2017, hariki2018,werner2022} we assume a near-exponential decay for the core hole, corresponding to a Lorentzian broadening of the XAS lines. In the real-time Keldysh formalism, the exponential decay of the core hole state can be introduced by attaching a wide-band particle reservoir to the core level, described by a core hybridization function $\Delta_c(t,t')$ and an action analogous to $S_{\Delta}$. As in Ref.~\onlinecite{werner2022}, we will use a Gaussian density of states for the core hybridization function $\Delta_c(\omega)=(\Gamma/\pi) \,e^{-(\omega/2W_c)^2}$, which yields a simple analytic form for $\Delta_c(t,t')$, and gives an exponential core-hole decay with a time constant $\Gamma^{-1}$  if the bandwidth $W_c$ is large.

\section{Minimal model}
\label{Sec:XAS_har}

The screened interaction  can be represented as arising from the coupling of the electron density to a bosonic field (Hubbard-Stratonovich field), whose spectrum is related to the charge fluctuations in the rest of the lattice. Before discussing the numerical results for charge-transfer insulators, we therefore consider a simple model which contains only a single harmonic screening mode. This model allows to illustrate the effect of the electron-boson coupling  on the XAS signal on a qualitative level. The corresponding Hamiltonian reads
\begin{align} \begin{split}
	H = &\,\epsilon_d n_d + \epsilon_c n_c + 
	U n_{d\uparrow} n_{d\downarrow} 	
	- \sqrt{2}\gamma \,\overline n_d X + \Omega_0 b^\dagger b,
	\label{eq:H_harm-screen}
\end{split}\end{align}
where $b^{(\dagger)}$ annihilates (creates) a screening excitation with energy $\Omega_0$, and $\gamma$ describes the coupling between the electron density fluctuations 
$\overline n_d=n_{d\uparrow} +n_{d\downarrow}-1 $ and the displacement $X=(b+b^{\dagger})/\sqrt{2}$ of the screening mode. The core and valence energy levels are $\epsilon_c$ and $\epsilon_d$, $U$ is the Hubbard repulsion in the valence orbital, and we choose $\epsilon_d=-U/2$ to enforce particle-hole symmetry. Any core-valence interaction is neglected as it will only produce an additional static energy shift on the absorption signal, because the model is purely local and the total electron density is preserved. In the initial state before the X-ray probe, the system is in the ground state with a single $d$-electron, and no screening mode is present.

\begin{figure}[t]
\centerline{\includegraphics[width=\linewidth]{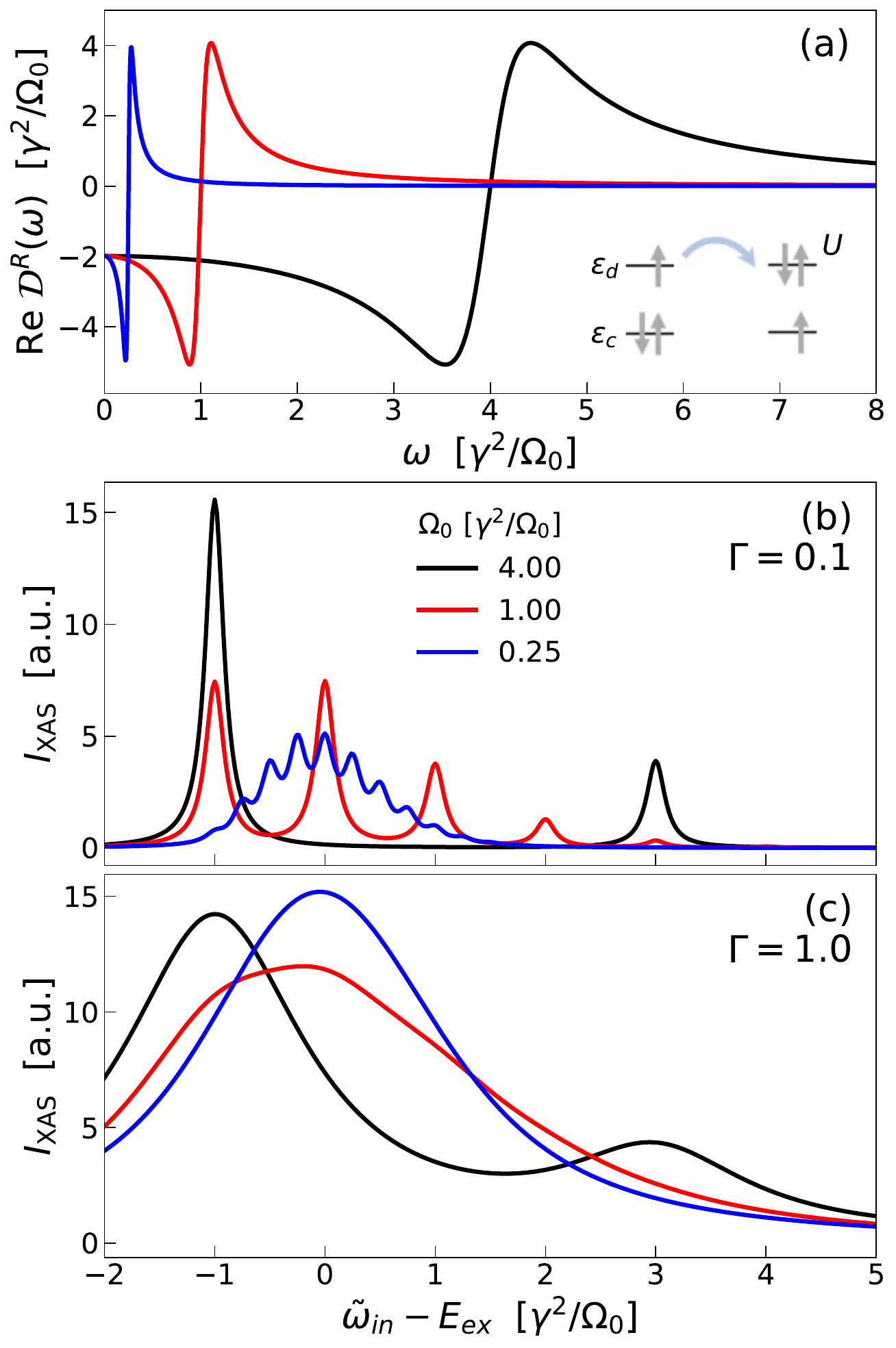}}
\caption{ (a) Real part of the effective retarded interaction \mbox{$\mathcal{D}^R(\omega+i0^+)$}  [Eq.~\eqref{Uretw}]  for $\Omega_0=0.25,1,4$. All energies in this figure are measured in units of $\gamma^2/\Omega_0$. Divergences at $\Omega_0$ have been smoothened for a better visualization. The inset shows the X-ray absorption process. In addition to the doublon, for $\gamma\neq 0$, the final state can contain an arbitrary number $m$ of quanta in the screening mode. 
(b) and (c) XAS signal of the minimal model, Eq.~\eqref{eq:xas_harm-screen}. The inverse core-hole lifetime is $\Gamma=0.1$ (b) and $1.0$ (c).}
\label{fig:harm-screen}
\end{figure}

The coupling between X-ray excited electrons and the screening mode gives rise to an additional, boson-mediated electronic interaction $\mathcal D$ on the $d$-orbital, which is obtained by integrating out the bosonic degrees of freedom in Eq.~\eqref{eq:H_harm-screen}. This yields a contribution to the action analogous to Eq.~\eqref{envU},  
$S_\mathcal{D}=-\frac{1}{2} \,\bar n_d\ast  \mathcal{D}\ast \bar n_d$, where the interaction
\begin{align} 
 \mathcal D(t,t')&= 2\gamma^2  \, D_0(t,t'),
 \label{Uret}
 \end{align}
is determined by the boson propagator $D_0(t,t')=-i\langle  T_{\mathcal C} X(t) X(t')\rangle$. In equilibrium, the retarded propagator becomes $D^R_0(\omega)=\Omega_0(\omega^2-\Omega_0^2)^{-1}$, which results in the boson-mediated interaction
\begin{align} 
\label{Uretw}
 \mathcal D^R(\omega)&= \frac{2\Omega_0\gamma^2}{\omega^2-\Omega_0^2}.
 \end{align}
 The real part of $ \mathcal D^R(\omega)$ is shown in Fig.~\ref{fig:harm-screen}(a). The induced interaction approaches the value 
\begin{align} 
\label{delU}
 \Delta U=-2\gamma^2/\Omega_0
 \end{align}
at low frequencies ($\omega\rightarrow 0$), and becomes zero for $\omega\gg \Omega_0$. If the energy of the screening mode is larger than all relevant energy scales in the model (anti-adiabatic limit), screening mode excitations are suppressed and one can approximate $\mathcal D^R(\omega)$ by its static limit $\Delta U$. In this case the dominant effect of screening is to modify the Hubbard interaction $U$ 
into the screened interaction $U+\Delta U$.
This analysis can be extended to investigate screening effects on XAS in more complex materials, where relevant screening modes have finite energy, as will be the case for the charge transfer insulator studied in Sec.~\ref{sec:results}.

Now we turn to the evaluation of the XAS signal in this minimal model. In the following, the photon energy is measured relative to the core level, $\omega_\text{in}=\tilde \omega_\text{in} +|\epsilon_c|$.
For $\gamma=0$, there is a single X-ray absorption line at the energy corresponding to the transition shown in the inset of Fig.~\ref{fig:harm-screen}(a). This main absorption process corresponds to $\tilde \omega_\text{in} = U+\epsilon_d= \frac{U}{2} \equiv E_{\text{ex}}$, the excitonic resonance. For $\gamma\neq 0$, additional excitations of the screening mode alter the XAS spectrum and we evaluate the XAS signal \eqref{eq:I_XAS_IMP1} analytically, employing a Lang-Firsov transformation (details on the derivation can be found in Appendix \ref{app:derivation_xas_minimal-model}). For a long probe pulse ($s(t)\to 1$) one obtains
\begin{align}
I_{\mathrm{XAS}}
\propto 
\sum_{m=0}^\infty  
\frac{2\Gamma \cdot e^{-\gamma^2/\Omega_0^2}(\gamma/\Omega_0)^{2m}/m! }
{( \tilde \omega_\text{in}-E_{\text{ex}}+\gamma^2/\Omega_0-\Omega_0 m) ^2 + \Gamma^2}.
		\label{eq:xas_harm-screen}
\end{align}
The spectrum shows an infinite series of satellite peaks, corresponding to the excitation of one doublon and  $m$ quanta $\Omega_0$ of the screening mode, each with a Lorentzian broadening due to the core-hole lifetime $1/\Gamma$. These boson satellites are  analogous to sidebands in the optical absorption of molecules with well-defined vibrational modes. The lowest peak $(m=0)$ is centered at the bare transition energy $E_{\text{ex}}$, shifted by the value  $\Delta U/2=-\gamma^2/\Omega_0$ due to the retarded interaction \eqref{delU}. The weights of the satellites follow a Poisson distribution with mean/standard~deviation $\gamma^2/\Omega_0^2$.

The XAS signal \eqref{eq:xas_harm-screen} for various parameters $\gamma,\Omega_0$ is shown in 
Fig.~\ref{fig:harm-screen}(b)-(c), with $\gamma^2/\Omega_0$ as the unit of energy. One can identify two regimes, with  distinct effects of the bosonic mode on the absorption spectrum: In the anti-adiabatic limit, $\Omega_0 \gg \gamma$ ($\Omega_0=4$ in the figure), the dominant effect is the aforementioned redshift of the 
main peak, along with weak satellites at 
an offset $m\Omega_0$, i.e.~far away from the excitonic resonance $E_{\text{ex}}$. In the adiabatic limit $\Omega_0\ll \gamma$, the dominant peak is no longer at $m=0$. Furthermore, due to the core-hole decay, the individual satellites broaden and merge 
into a single asymmetric line shape if $\Omega_0 \lesssim \Gamma$ (see Fig.~\ref{fig:harm-screen}c). Thus, the superposition of many boson satellites appears as an additional broadening of the XAS absorption spectrum into an asymmetric peak. Because the weights of the satellites follow a Poisson distribution, the mean energy is at $\Omega_0 \cdot \gamma^2/\Omega_0^2$, above the redshifted main absorption line, such that the mean position of this asymmetric line 
remains centered at the unscreened $\tilde\omega_\text{in}=E_{\text{ex}}$. For finite $\Gamma$ and in the ultra-adiabatic regime $\Omega_0\to0$ ($\Omega_0 =0.25$ in the figure), the XAS signal approaches again a symmetric (Gaussian) form centered around $\tilde\omega_{\text{in}}=E_{\text{ex}}$.

With these insights, it will be interesting to see whether the effect of the screening in typical charge transfer insulators also manifests itself in a modified lineshape. In addition to this qualitative discussion, the minimal model can serve as a benchmark for the approximate NCA impurity solver used in the remainder of the manuscript (see also the comment in Sec.~\ref{XASGWEDMFT}). It turns out that the NCA can reproduce the exact result \eqref{eq:xas_harm-screen} reasonably well in the adiabatic $\Omega_0\ll\gamma$ and anti-adiabatic $\Omega_0\gg\gamma$ regime, while for intermediate boson frequencies significant deviations can occur (see Appendix~\ref{app:nca_harm-screen} for details). 

\section{XAS in charge-transfer insulators}
\label{sec:results}

\subsection{Model}

We consider a two-dimensional three-band Emery model~\cite{emery1987} on the square lattice with strongly interacting Cu orbitals, marked with an index $d$, and oxygen orbitals, marked with indices $p_x$ and $p_y$, as relevant for cuprate superconductors. The total Hamiltonian consist of the terms
\begin{align}
  &H_{dp}=H_\text{e}+H_\text{kin}+H_\text{int}, \label{Eq:dp} \\
  &H_\text{e}= \epsilon_d \sum_i n_{id} +\epsilon_p \sum_i (n_{ip_x}+n_{ip_y}) , \nonumber\\
  &H_\text{kin}=\sum_{ij\sigma}\sum_{\alpha,\beta} t_{ij}^{\alpha\beta} d_{i\alpha\sigma}^{\dagger} d_{j\beta\sigma}, \nonumber
\end{align}
where $d_{i\alpha\sigma}$ is the annihilation operator for unit cell $i$, orbital $\alpha\in\{d,p_x,p_y\}$, and spin $\sigma$; $n_{i\alpha}=n_{i\alpha,\uparrow}+n_{i\alpha,\downarrow}$ is the corresponding density operator. 
The interaction is a density-density interaction of the general form \eqref{Hintgeneral}, with $v_{i-j}^{\alpha\beta}=U$ if $i=j$ and $(\alpha;\beta)=(d\!\uparrow;d\!\downarrow)$ or $(d\!\downarrow;d\!\uparrow)$, and $v_{i-j}^{\alpha\beta}=U_{dp}$, when $(i\alpha)$ and $(j\beta)$ are nearest-neighbor $d$ and $p$ orbitals. The on-site energy of the $d$ orbital is given by $\epsilon_d$, and the crystal field splitting between the $d$ and $p$ orbitals $\Delta_{dp}=\epsilon_p-\epsilon_d$ is chosen such that the $p$ orbitals are completely filled, while the $d$ orbitals are singly occupied. The hopping between nearest neighbor $d$ and $p_x$, $p_y$ orbitals is denoted by $t_{dp}$ and between the $p_x$ and $p_y$ orbitals by $t_{pp}$.

In all calculations, we will fix the parameters to values relevant for La$_2$CuO$_4$, i.e., $U=5.0$ eV, $U_{dp}=2.0$ eV, $t_{dp}=0.5$ eV, $t_{dd}=-0.1$ eV, $t_{pp}=0.15$ eV and $\Delta_{pd}=-3.5$ eV;  the inverse temperature is $\beta=5.0$~eV$^{-1}$. As the local interactions on the $d$ orbital are strongest, the EDMFT description is restricted to the $d$ orbitals only, while the $p$ orbitals are treated within the GW approximation. The same model has been studied within GW+EDMFT in Refs.~\onlinecite{golez2019a,golez2019} in order to predict the photo-induced changes of the spectral function and the optical conductivity. Here we extend these studies  to the XAS signal. 

\subsection{Spectral function and screened interaction}

\begin{figure}[t]
\centerline{\includegraphics[width=\columnwidth]{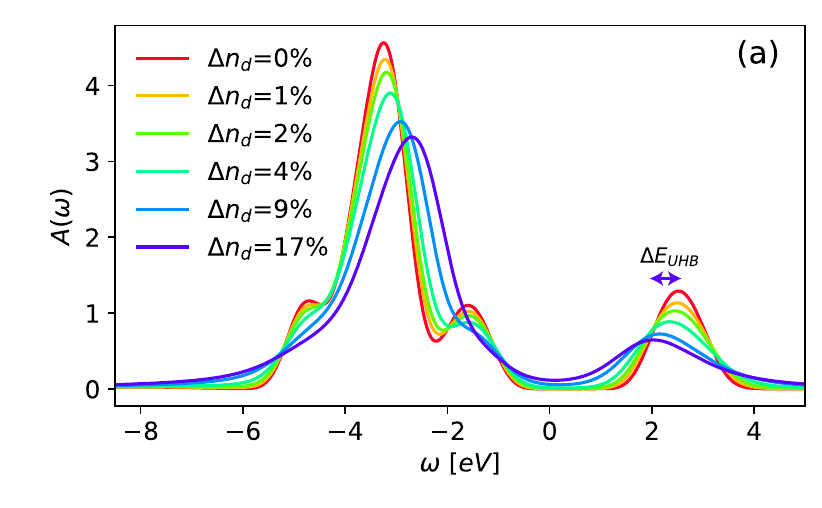}}
\centerline{\includegraphics[width=\columnwidth]{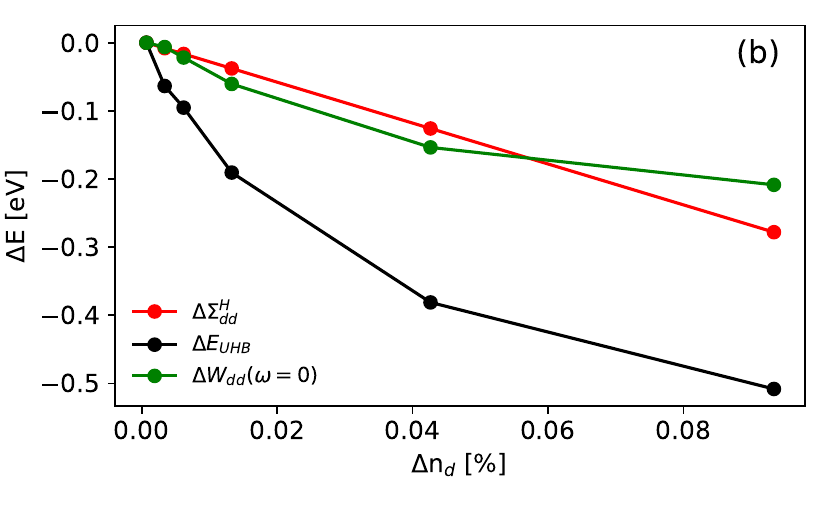}}
\caption{(a)~Local spectral function in equilibrium~(red line) and $t=12$ fs after the photo-excitation  for various excitation densities  $\Delta n_d$.  $\Delta E_{\text{UHB}}$ indicates the shift of the upper Hubbard band upon photo-doping. (b) Hartree shift $\Delta\Sigma_{dd}^H$ [Eq.~\eqref{Hartree_shuft}], shift of the upper Hubbard band $\Delta E_{\text{UHB}}$, and change  in the fully screened interaction  $\Delta W_{dd}(0)/2$  (see main text) as a function of photo-doping $\Delta n_d$.}
\label{Fig:Spectral}
\end{figure}

For completeness, we will first reproduce from Ref.~\onlinecite{golez2019a} the spectral properties of the $d$-$p$ subspace in equilibrium and after the photo-excitation. The spectral function is obtained from the two-time Green's functions as $A(\omega,t)=-\frac{1}{\pi}\int_t^{t+t_{\text{cut}}} dt' e^{\I\omega (t'-t)} G^R(t',t)$ with a cutoff time $t_{\text{cut}}=8$ fs both for the system with and without excitation. In Fig.~\ref{Fig:Spectral}(a), we show the spectral function in equilibrium and after photo-excitation. The almost fully occupied $p$ orbital is strongly hybridized with the lower Hubbard band of the $d$ orbital, forming three distinct features:  The bonding Zhang-Rice band around $\omega=-1$~eV, a band of predominantly $p$ character around $\omega=-3$~eV, and the anti-bonding band at the highest binding energies. The unoccupied part of the spectrum is mainly composed of the $d$ orbital upper Hubbard band with a very small admixture of the $p$ orbital around $\omega=3$~eV. In the atomic limit, the position of the upper Hubbard band is  
\begin{align}
E_{\text{UHB}}=U+\epsilon_d.
\label{euhb}
\end{align}

To model the photo-excitation, the action of an electric field $\vec{E}(t)$ with  polarization along the (11) direction is simulated. As found in Ref.~\onlinecite{golez2019a}, the excitation is followed by  rapid intra-band relaxation processes on the timescale of a few femtoseconds, after which the system remains for a longer time  in a non-thermal photo-doped state, whose properties are mainly controlled by the excitation density (photo-doping). The latter will be defined as the change $\Delta n_d$ in the $d$ occupancy. Because there are two $p$ orbitals and one $d$ orbital in each unit cell, there is a corresponding change $\Delta n_p=\tfrac12\Delta n_d$ in the occupation per $p$ orbital ($\Delta n_{p}\equiv \Delta n_{p_x}=\Delta n_{p_y}$). In the present paper, we focus on the analysis of this photo-doped state, measured at the longest accessible time $t=12$~fs, rather than on the excitation process itself. 
Specifically, the field amplitude has a time profile 
\begin{align}
E(t) = E_0 e^{-4.6(t-t_0)^2/t_0^2} \sin(\Omega(t-t_0)),
\end{align}
with a frequency $\Omega$ that is resonant with the charge transfer transition between the band with predominant $p$ character and the upper Hubbard band ($\Omega=6$ eV), and a duration $t_0 = 2\pi n/\Omega$ such that the envelope accommodates $n = 4$ cycles. The field is coupled to the tight-binding model \eqref{Eq:dp} using  the Peierls substitution and a dipolar matrix element of magnitude $0.3 ea$ ($e$ is the electronic charge and $a$ the lattice constant), see Ref.~\onlinecite{golez2019a} for details. The amplitude  $E_0$ of the electric field $\vec E(t)$ is adjusted to fix the amount of photo-doping. 

After the photo-excitation, the spectrum is modified in two characteristic ways, see Fig.~\ref{Fig:Spectral}(a). First, the bands are shifted toward the chemical potential, so that the bandgap is reduced. The analysis in Ref.~\onlinecite{golez2019a} showed that the photo-induced bandgap renormalization originates from two effects: The redistribution of charges between the $d$ and $p$ orbitals leads to a mean-field potential on the $d$ band,
\begin{align}
\Delta \Sigma^{H}_{dd}=\frac{U}{2}\Delta n_d - 4U_{dp}\Delta n_p
=
\Big(\frac{U}{2} - 2U_{dp}\Big)\Delta n_d,
\label{Hartree_shuft}
\end{align}
where the first factor $1/2$ is because the on-site Hubbard interaction acts only between electrons of different spin, and the factor $4$ counts the nearest-neighbor $p$ orbitals of a $d$ orbital. In a simpler Hartree Fock + DMFT (HF+DMFT) simulation, this mean field potential approximately explains the shift of the upper Hubbard band (Hartree shifts)~\cite{golez2019,golez2019a}. In the GW+EDMFT simulation, the shift $\Delta E_{\text{UHB}}$ of the upper Hubbard band is larger, indicating the effect of dynamical screening of the on-site Coulomb interaction. Both the Hartree shift and $\Delta E_{\text{UHB}}$ increase in magnitude with photo-doping, see Fig.~\ref{Fig:Spectral}(b). The second observation is that all bands are substantially broadened, which has been interpreted as a consequence of scattering of $d$ electrons on charge fluctuations \cite{golez2019a,golez2019,golez2022}. At the largest excitations strengths, the broadening is so large that the fine structure in the occupied part of the spectrum is completely washed out.

\begin{figure}[t]
\centerline{\includegraphics[width=\columnwidth]{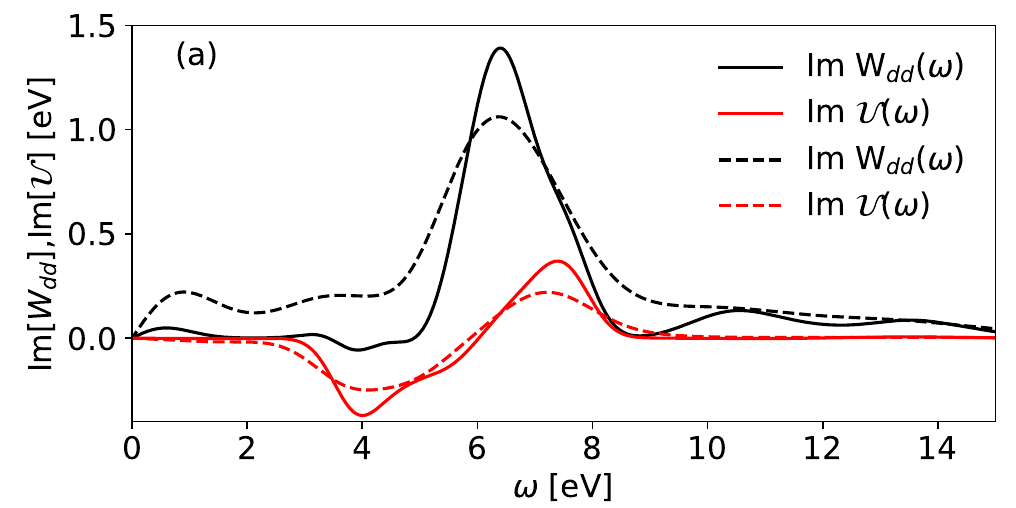}}
\centerline{\includegraphics[width=\columnwidth]{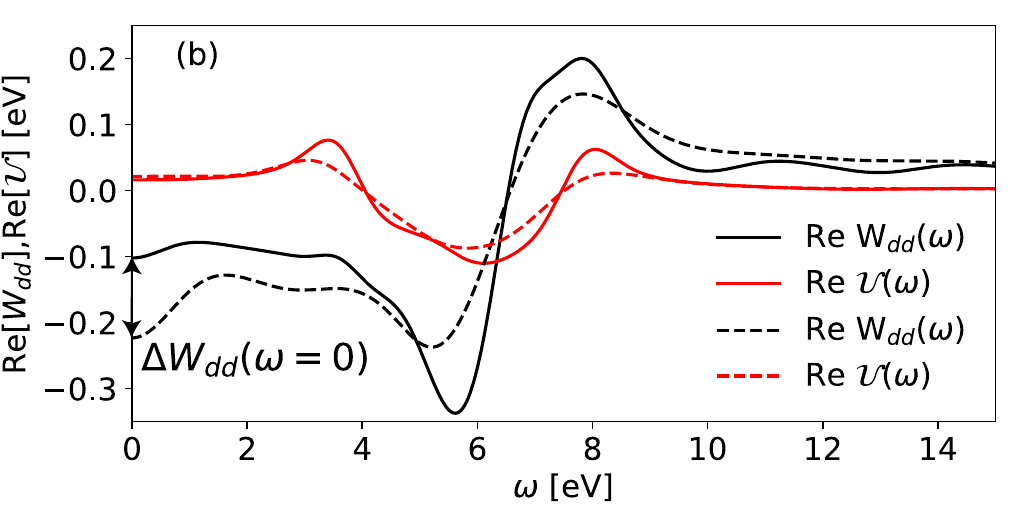}}
\caption{
Imaginary (a) and real part (b) of the fully screened interaction $W_{dd}(\omega)$ and the effective impurity interaction $\mathcal{U}(\omega)$. Full lines show the equilibrium result, while dashed lines correspond to a photo-doped state with $\Delta n_d=4\%$ measured $t=12$~fs after the photo-excitation. $\Delta W_{dd}(\omega=0)$ marks the renormalization of the static interaction presented in Fig.~\ref{Fig:Spectral}.}
\label{Fig:Spectral_W}
\end{figure}

Along with the photo-induced changes in the spectrum also the screening environment is substantially modified. In  Fig.~\ref{Fig:Spectral_W},  we show the local fully screened interaction $W_{dd}(\omega,t)=\int_t^{t+t_{\text{cut}}} dt' e^{\I\omega (t'-t)} W^R_{dd}(t',t)$ on the $d$-orbitals. In equilibrium, the spectrum $-\text{Im}W_{dd}(\omega)$ [Fig.~\ref{Fig:Spectral_W}(a)] shows a pronounced peak at the energy $\omega\approx7$~eV, 
because screening processes are mainly associated with virtual charge fluctuations between the $p$ band and the upper Hubbard band.  After the photo-excitation, additional spectral weight appears in $W_{dd}$ below $2$ eV, which is related to  charge fluctuations of photo-doped doublons in the upper Hubbard band and holes in the  Zhang-Rice band.  Consequently, after the photo-doping, the real part $\text{Re}W_{dd}(\omega)$ 
[Fig.~\ref{Fig:Spectral_W}(a)] is suppressed below $\omega=1$~eV by an amount $\Delta W_{dd}(0)=\text{Re}W_{dd}(\omega=0)-\text{Re}W_{dd}(\omega=0)|_{\Delta n_d=0}$, which is another manifestation of dynamical screening. Assuming that the reduction of the effective interaction $\Delta W_{dd}(\omega=0)$ symmetrically modifies the gap size, 
we observe that $\Delta W_{dd}(\omega=0)/2$ and the Hartree shift each make a comparable contribution to the shift of the upper Hubbard band as the photodoping is increased [Fig.~\ref{Fig:Spectral}(b)].

The effective impurity interaction  Im$\,\mathcal{U}(\omega)$ has a dip and peak structure with negative weight at positive frequencies already in equilibrium (Fig.~\ref{Fig:Spectral_W}(a)). The fact that $\text{Im}\,\mathcal{U}(\omega)$ can become negative in GW+EDMFT calculations is well known~\cite{golez2017,boehnke2016,lee2017,vucicevic2018}; while it does not necessarily imply a non-causal behavior in observable quantities, there are alternative formalisms of the self-consistency which enforce a positive spectral function of $\mathcal{U}(\omega)$~\cite{backes2022,chen2022}.  Additionally, the combination of the NCA impurity solver with the GW+EDMFT can cause slight violations of causality \cite{golez2019}. Because $\mathcal{U}(\omega)$ is an auxiliary quantity within the present GW+EDMFT formalism, we will  rely on observable quantities such as spectra and the fully screened interaction in the following discussion.

\subsection{XAS - Atomic limit}

We will focus on the XAS signal of the strongly-interacting $d$ orbital, with a single core level, 
\begin{align}
\label{Hc_}
  H_{c}=\epsilon_c n_{c},
\end{align}
at energy $\epsilon_c<0$. In addition, there is typically a strong interaction between the core hole and the valence electrons,
\begin{align}
\label{Hcd_}
	H_{cd}=U_{cd} (n_{c}-2) n_{d},
\end{align}
defined such that the contribution vanishes for a system without core hole. Unless otherwise stated, we will consider the empirical relation $U_{cd} \approx 1.3 U_{dd} =6.5$ eV, which has proven accurate for a broad range of transition metal oxides~\cite{laan1986,hariki2017,hariki2018,zaanen1986}. The XAS signal in Eq.~\eqref{eq:I_XAS_IMP1} is evaluated with a Gaussian probe envelope $s(t)$ of duration  $\sigma=3.6$ fs. 
From now on, we will state the X-ray energies relative to the core energy, $\tilde \omega_\text{in} = \omega_\text{in}-|\epsilon_c|$. 

\begin{figure}[t]
\includegraphics[width=\columnwidth]{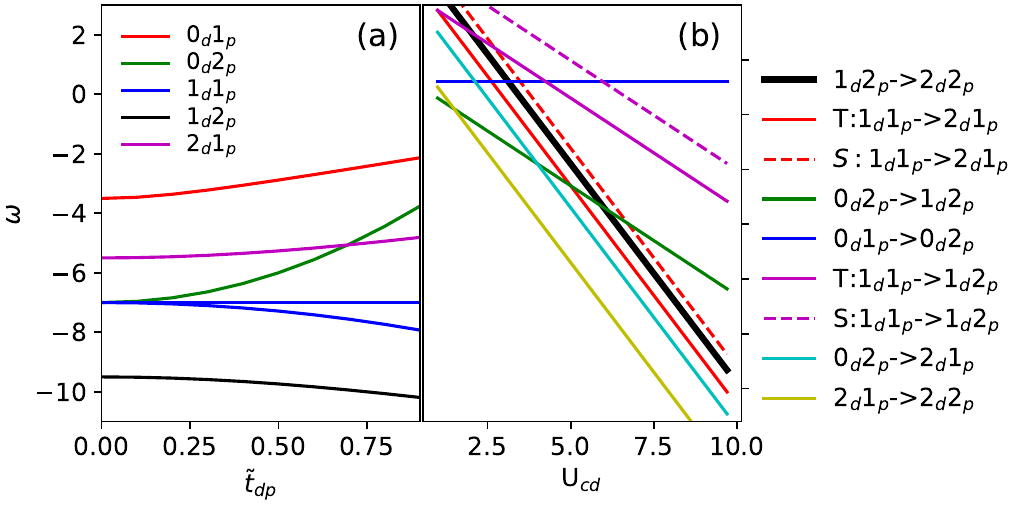}
\caption{
(a) Multiplet states of the cluster model \eqref{cluster_}, as the inter-cluster hopping is turned on ($t_{pp}$ is scaled along with $t_{pd}$).
Groups of multiplets are labelled according to their dominant atomic configuration $m_dn_p$. Explicit expressions for the states are given in the App.~\ref{app:cluster}. (b) X-ray absorption will occur at $\tilde \omega_\text{in}=E_{a\to b}$, where the energies $E_{a\to b}$ are shown as a function of $U_{cd}$.  The color and labelling of the lines only indicates from which group of multiplets the initial and final state of the X-ray transition is taken. The transition $1_d2_p\to\underline{2}_d2_p$ from the ground state configuration is shown in bold. The value for $U_{cd}$ used in the main text is $U_{cd}=6.5$ and $\tilde t_{dp}=1.0$.}.
\label{Fig:atomic_limit}
\end{figure}

Because the final states of the X-ray absorption process are often localized due to the strong core-valence interaction, it will be useful to support the  interpretation of the numerical XAS spectra by a cluster model. Here, we 
introduce this cluster model and its multiplet states for later reference. We can start from a CuO$_{4}$ plaquette, i.e., one $d$-orbital in Eq.~\eqref{Eq:dp} with four surrounding $p$ orbitals. Due to to $C_4$ symmetry, only the symmetric~(B$_{1g}$) combination of the four oxygen orbitals will hybridize with the copper $d_{x^2-y^2}$ orbital~\cite{zhang1988, zannen1988, ramsak1989, feiner1996, Jefferson1992}.  
This leads to a model with one $d$ orbital and the symmetric $p$ orbital,
\begin{align}
\label{cluster_}
H= H_{\text{at}} + H_{c} + H_{cd} - \tilde t_{pd}\sum_{\sigma} ( d_{\sigma}^\dagger p_{\sigma} + h.c.) ,
\end{align}
where 
\begin{align}
H_{\text{at}}=Un_{d\uparrow} n_{d\downarrow}+
\tfrac12 U_{dp} n_d n_p+\epsilon_d  n_d +\tilde \epsilon_pn_p, 
\end{align}
and the core contribution is given by Eqs.~\eqref{Hc_} and \eqref{Hcd_}.
The parameters $\tilde \epsilon_p=\epsilon_p-1.5 t_{pp}$ and $\tilde t_{dp}=2 t_{dp}$
are shifted with respect to the lattice Hamiltonian \eqref{Eq:dp} due to the mapping on the symmetrized $p$ orbital. In principle, orthogonalization of orbitals on neighboring plaquettes would lead to a further small renormalization  $\tilde\epsilon_p \rightarrow \epsilon_p-1.45 t_{pp}$, $U_{dp}\rightarrow 0.92 U_{dp}$ and $\tilde t_{dp}\rightarrow 1.92 t_{dp}$~\cite{feiner1996}, but because we use the cluster model only for a qualitative comparison to the numerical data, we will neglect these shifts in the following. 

In the atomic limit $t_{pd}=0$, multiplet states are simply given by integer occupancies, and will be denoted by $m_{d}n_p$, where $m~(n)\in \{0,1,2\}$ is the number of  electrons on the $d$~($p$) orbital. An underline as in $\underline{m}_{d}n_p$ will denote the state with an additional core hole. We will keep the same notation also for states with nonzero $t_{dp}.$ The initial state is therefore $1_d2_p$, and the XAS transition corresponds to $1_d2_p$ $\to$ $\underline{2}_d2_p$. For non-vanishing $t_{dp}$, one finds small shifts and splittings of the levels on the order O($t_{dp}^2/U_{dd}$,$t_{dp}^2/\Delta_{dp}$),  see Fig.~\ref{Fig:atomic_limit}(a). For completeness, a detailed analysis of the states in such a cluster is presented in App.~\ref{app:cluster}. We will refer to the corresponding X-ray transition energies  (Fig.~\ref{Fig:atomic_limit}(b)) in the following analysis of the numerical GW+EDMFT results.

\begin{figure}[t]
\centerline{\includegraphics[width=\linewidth]{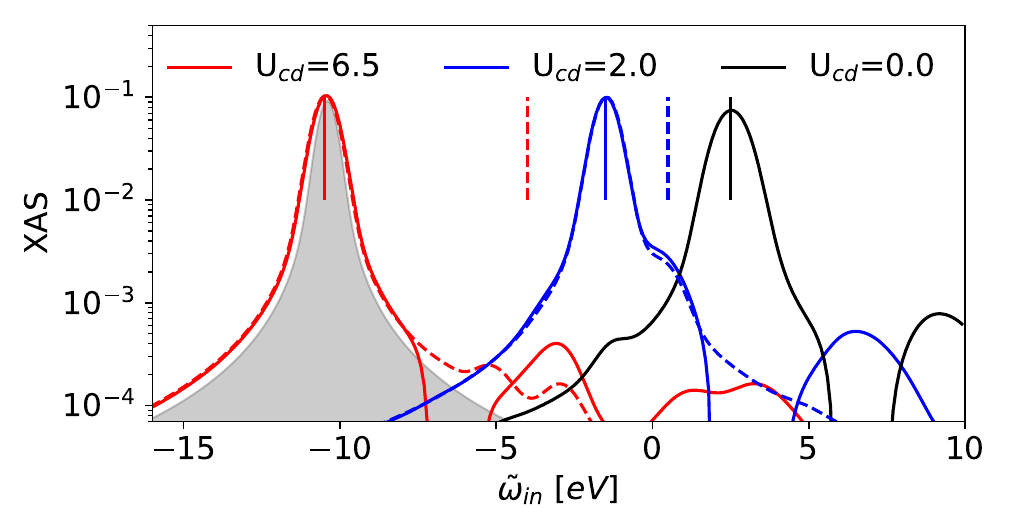}}
\caption{
Equilibrium XAS as a function of the energy $\tilde \omega_{\text{in}}$ for different values of the core-valence interaction $U_{cd}$ calculated within GW+EDMFT. Dashed lines correspond to a Hartree-Fock + DMFT simulation, where the effect of dynamical screening is missing. The full vertical lines present the atomic estimate $E_\text{ex}$ of the exciton transition $1_d2_p\rightarrow \underline{2}_d2_p$ 
[Eq.~\eqref{exbare}], and the dashed vertical lines mark the position  $E_\text{ex}+U_{cd}$ of the continuum. The shaded region around $\tilde \omega_{\text{in}}=-10.5$ is a convolution of the probe pulse and the lifetime of the core hole rescaled to the height of the dominant peak, serving as an estimate of its lifetime.}
\label{Fig:xas_eq}
\end{figure}

\subsection{XAS - Equilibrium system}
\label{xasequi}

In equilibrium, the $d$ orbital is half-filled and the $p$ orbital is completely filled. In the atomic limit ($t_{pd}=0$),  the dominant state is therefore $1_d2_p$, and the X-ray absorption corresponds to the transition $1_d2_p\rightarrow \underline{2}_d2_p$ at $\tilde \omega_\text{in}=E_\text{ex}$,  with 
\begin{align}
\label{exbare}
E_{ex}=U+\epsilon_d-2U_{cd} = E_{\text{UHB}}-2U_{cd}.
\end{align}
(In the extended cluster model \eqref{cluster_}, with $\tilde t_{pd}=1.0$, the transition energy is only slightly shifted by $\Delta E=0.25$ eV.) Figure~\ref{Fig:xas_eq} shows the XAS spectrum obtained from the GW+EDMFT simulation in equilibrium. One indeed observes a main peak around energy  $\tilde \omega_\text{in}=E_\text{ex}=-10.5$~eV.  The final state $\underline{2}_d2_p$ is a local bound state (exciton), because the transfer of one $d$ electron of the local state $\underline{2}_d2_p$ to the continuum (i.e., to a different lattice site, leaving the configuration $\underline{1}_d2_p$ at the position of the core hole) would cost 
an energy $U_{cd}$ which exceeds the energy gain from delocalization. The line-shape of the main peak therefore well matches the convolution of a Lorentzian with inverse width $1/\Gamma=12$ fs due to the core hole decay and a Gaussian with width  $\sigma=4$ fs due to the finite probe duration, see the dashed region in Fig.~\ref{Fig:xas_eq} and the discussion in Ref.~\onlinecite{werner2022}. The missing data points around $\omega=-6$~eV and $-1$~eV correspond to regions of slightly negative spectral weight, which is a numerical artifact related to the non-causal effects discussed above and to the limited numerical accuracy. 

Due to charge fluctuations in the initial state, the XAS transition can also lead to a final state in the continuum (local configuration $\underline{1}_p2_p$ with an additional $d$ electron in the rest of the lattice). In fact, we find a weak satellite to the main exciton at energy $\tilde \omega_\text{in}=E_\text{ex}+U_{cd}\approx -3$~eV (see dashed line).  An alternative interpretation of this satellite could be a boson satellite due to the coupling to dynamic charge fluctuations. It is therefore instructive to compare the  GW+EDMFT calculation with the HF+DMFT result, which does not include screening, see dashed-dotted line in Fig.~\ref{Fig:xas_eq}. The comparison shows that the satellite signal, although overall weak,  is  enhanced if screening is taken into account, which shows that this peak is at least in part originating from charge fluctuations. 

While in  transition metal oxides there is typically a fixed relation between $U_{dd}$ and $U_{cd}$ \cite{laan1986,hariki2017,hariki2018}, it is instructive to separate 
the two interactions by artificially adjusting $U_{cd},$ see Fig.~\ref{Fig:xas_eq}.  For  $U_{cd}=2$ eV, the continuum forms a broad shoulder on top of the dominant resonance and the sidepeak due to bosonic excitations is now clearly present around $\omega-\omega_{\text{in}}=6$ eV. For $U_{cd}=0$, there is no formation of the 
exciton, but the sidepeak is clearly visible.

The fact that the bosonic satellite is rather weak for the present case can be rationalized with the minimal model of Sec.~\ref{Sec:XAS_har}, 
although it is clear that a comparison of the minimal model with the GW+EDMFT impurity model should not be over-interpreted. The two models can be related by the induced interaction, but one needs to take into account one subtle point: Because the charge susceptibility of an isolated site vanishes, the fully screened interaction $W_{dd}$ and the induced interaction \eqref{Uret} are identical in the minimal model  \eqref{eq:H_harm-screen}. In contrast, 
$W_{\text{imp}}$ and $\mathcal{U}$ differ   in the  GW+EDMFT  impurity model, because the hybridization between the impurity and bath leads to charge fluctuations on the $d$ orbital in the impurity model.
The minimal model should be viewed as a toy model obtained from the GW+EDMFT impurity model after approximately eliminating the charge fluctuations from the environment, which will renormalize the fully screened interaction.    We therefore link the two models by comparing the fully screened interaction $W_{dd}$ to the induced interaction \eqref{Uret}, instead of the effective interaction $\mathcal{U}$. With this interpretation we match the parameters $\Omega_0$ and $2\gamma_0^2/\Omega_0$ in the minimal model with the position of the main peak in $\text{Im} W_{dd}(\omega)$ ($\Omega_0\approx7$~eV) and the reduction $U-\text{Re}W_{dd}(\omega=0)\approx 0.1$~eV, respectively.  This places the minimal model clearly in the anti-adiabatic regime, with $\gamma_0^2/\Omega_0^2 =0.01$, which explains the very weak nature of the bosonic satellites in the XAS spectrum (in the anti-adiabatic limit, the first satellite in Eq.~\eqref{eq:xas_harm-screen} is smaller than the main peak by a factor $\gamma_0^2/\Omega_0^2$).

\subsection{XAS - Photo-doped system}

\begin{figure}[t]
\centerline{\includegraphics[width=\columnwidth]{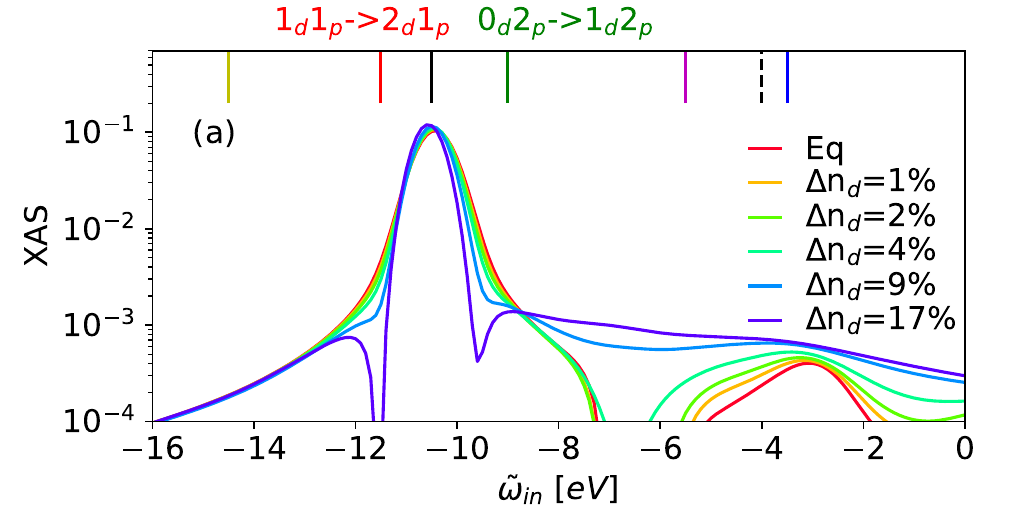}}
\centerline{\includegraphics[width=\columnwidth]{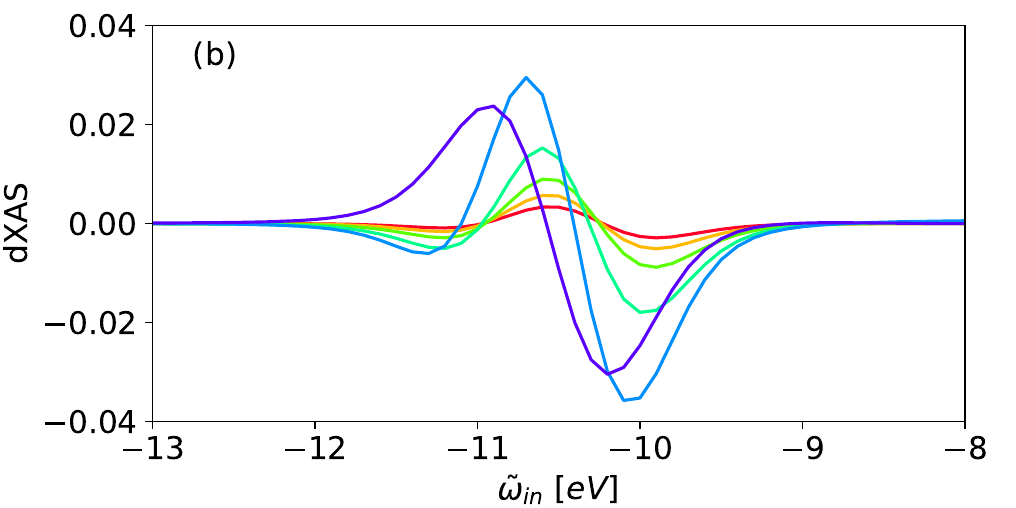}}
\caption{
(a)~XAS snapshot after the photoexcitation (pump-probe delay $12$ fs) for the indicated photodoping  $\Delta n_d$.  (b)~Difference of the XAS signal with respect to the  equilibrium signal~(dXAS). The vertical lines in panel (a) indicate the X-ray transitions originating from excited initial states in the $p$-$d$ cluster at $U_{cd}=6.5$~eV
and $\tilde t_{pd}=1.0$, using the same color code for the transitions as in Fig.~\ref{Fig:atomic_limit}(b). Selected transitions are labelled, such as the lowest transition from an initial state with a photo-doped $d$-hole ($0_d2_p$ $\to$  $\underline{1}_d2_p$) and with a photo-doped $p$-hole ($1_d1_p$ $\to$  $\underline{2}_d1_p$).}
\label{Fig:xas_U6}
\end{figure}

\begin{figure}[t]
\centerline{\includegraphics[width=\columnwidth]{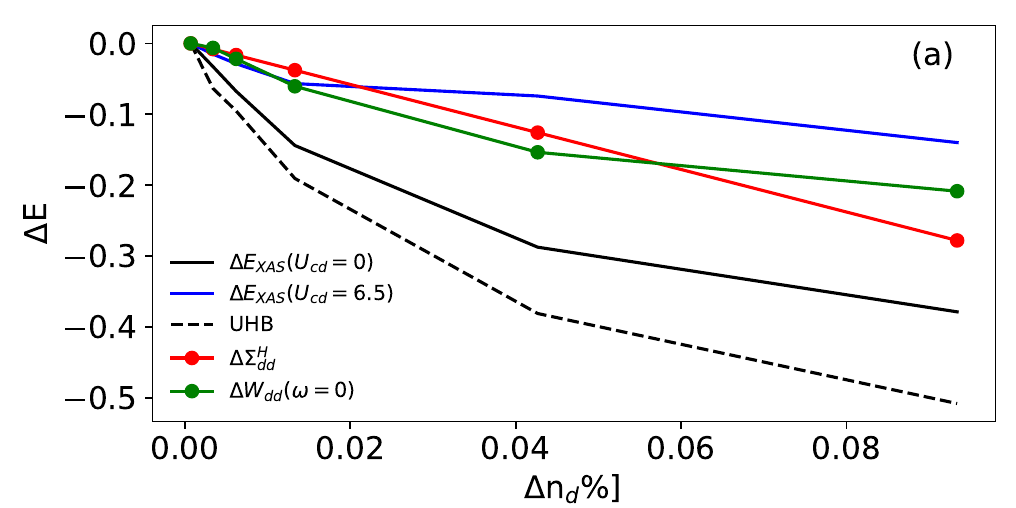}}
\centerline{\includegraphics[width=\columnwidth]{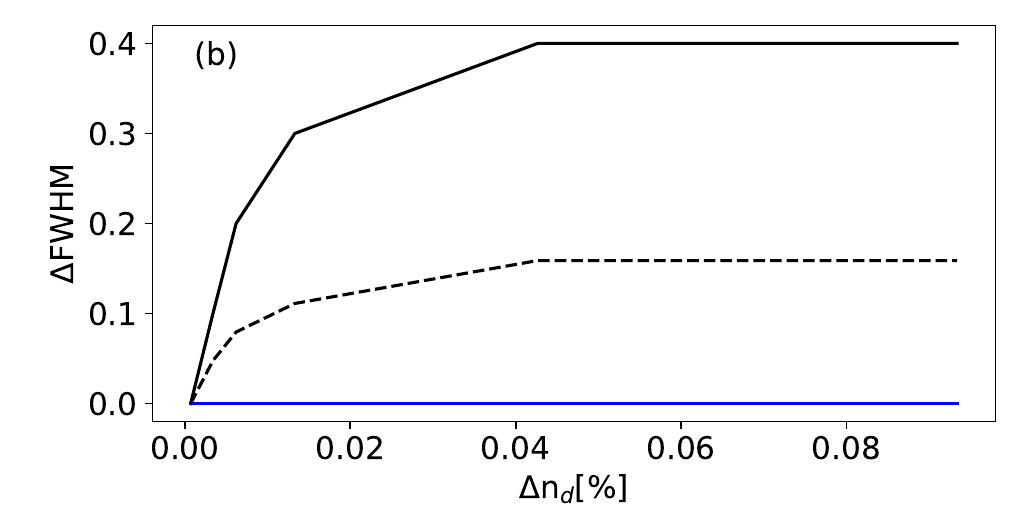}}
\caption{(a)~Photoinduced shift in the excitonic energy versus the photodoping for $U_{cd}=6.5$ eV and $U_{cd}=0$ eV, compared to the shift of the upper Hubbard band~(UHB), the Hartree shift $\Delta\Sigma^H_{dd}$ and the changes in the screening $\Delta W_{dd}/2$. (b)~Change in the linewidth (FWHM)  of the excitonic peak versus the photodoping $\Delta n$ (same labeling as in panel (a)). Here, the dashed line (UHB) indicates the change in the FWHM of the upper Hubbard band obtained from the local spectral function.}
\label{Fig:scaling}
\end{figure}

We will now analyze how the XAS signal is modified after a photo-excitation and what information can be extracted from these changes. We will focus on the physically relevant regime of strong core-valence interaction $U_{cd}=6.5$ eV and analyze the change in the spectrum as a function of the photodoping $\Delta n_d$, see Fig.~\ref{Fig:xas_U6}. The dominant change in the spectrum is a shift of the exciton to larger binding energies, see also Fig.~\ref{Fig:xas_U6}(b), which plots the difference with respect to equilibrium. Furthermore, additional spectral weight appears, mainly at smaller binding energies. Here, the dominant effect is the increase of the signal at the XAS continuum $\tilde \omega_\text{in}\approx -4$ in Fig.~\ref{Fig:xas_U6}, which is expected as the photo-doping increases the number of mobile charge carriers.

First, we focus on the photo-induced shift $\Delta E_\text{ex}$ of the main exciton line. 
The
maximum is determined by fitting a qubic spline and extracting the maxima from the interpolated curve. Figure~\ref{Fig:scaling}(a) shows the shift as a function of the photo-doping $\Delta n_d$. Since the position of the peak in the atomic analysis is given by Eq.~\eqref{exbare}, 
the peak provides a first measure for the change in $U$. 
However, a closer analysis shows that there are two important facts one should keep in mind regarding this interpretation:

(i) In analogy to the shift of the upper Hubbard band discussed in Fig.~\ref{Fig:Spectral}(b), part of the shift can be explained by the Hartree shift \eqref{Hartree_shuft}  of the $d$ level. The red line in Fig.~\ref{Fig:scaling}(a) shows the Hartree shift (same as in Fig.~\ref{Fig:Spectral}(b)). One can see that while the Hartree shift partially explains the renormalization of the exciton, the remaining shift nevertheless indicates a substantial contribution from dynamical screening and the renormalization of $U$.

(ii) With the shift of the XAS exciton and the upper Hubbard band, we now in principle have two independent experimental measures for the renormalization of the interaction, namely XAS and inverse photoemission spectroscopy~\cite{smith1988} or two-photon photoemission~\cite{gillmeister2020,petek1997}. Comparing the expressions for the two quantities in the atomic limit (Eqs.~\eqref{euhb} and \eqref{exbare}), one could expect that the two measures shift by the same amount if the effect of screening can be described by a simple renormalization of the Hubbard $U$. In contrast, by comparing $\Delta E_{\text{UHB}}$ and $\Delta E_{\text{XAS}}(U_{cd}=6.5)$ in Fig.~\ref{Fig:scaling}(b), we see that the shift of the upper Hubbard band after photo-doping is substantially larger than the shift of the XAS exciton. (Because for $U_{cd}=0$, XAS is identical to electron addition up to core-hole lifetime effects, $\Delta E_{\text{UHB}}$ closely matches the shift $\Delta E_{\text{XAS}}(U_{cd}=0)$ for a simulation with $U_{cd}=0$, shown in Fig.~\ref{Fig:scaling}(a).)  

In part, the difference between $\Delta E_{\text{UHB}}$  and  $\Delta E_{\text{ex}}$ is due to the fact that the change of the lineshape of the two spectral features upon photo-doping is substantially different (see Fig.~\ref{Fig:scaling}(b)): While a strong broadening of the upper Hubbard band is observed, the line-width of the exciton changes only weakly. This difference is expected, as the delocalized states in the UHB induce a change in the electronic spectral function due to their scattering with photo-doped holes, in contrast to the localized exciton. The asymmetric broadening of the upper Hubbard band can also influence the determination of its mean-position.
Beyond the different broadening of  the exciton and the upper Hubbard band, a possible reason for the difference between $\Delta E_{\text{UHB}}$  and  $\Delta E_{\text{ex}}$ is the  combined effect of screening and charge fluctuations on the level shifts. In the minimal atomic model \eqref{eq:H_harm-screen}, which does not include any charge fluctuations on the impurity site, the effect of the screening mode on $E_{\text{UHB}}$  and  $E_{\text{ex}}$ would indeed be identical. In the presence of the lattice, however, the final states with a doubly occupied impurity site are hybridized with unoccupied continuum states (in the GW+EDMFT impurity model, these appear as unoccupied spectral density in the hybridization function $\Delta$ around $\omega=E_{\text{UHB}}$). The hybridization shifts are smaller for the exciton, because continuum states are far off-resonant. Since the coupling to dynamic bosonic modes does not only modify the interaction, but also affects the tunnelling rate to the continuum, a different shift of $E_{\text{UHB}}$  and  $E_{\text{ex}}$ is indeed expected. 

Another issue to be discussed is the interaction between $p$ electrons and the core. So far, we have assumed that the core electrons interact only with the electrons at the same site, via the interaction $U_{cd}$, 
while for the valence orbitals, 
a nearest-neighbor interaction $U_{dp}$ is considered. More generally, we could introduce an oxygen-core interaction $U_{cp}$ as a free parameter in the Hamiltonian. Our assumption $U_{cp}=0$ is in line with the suggestion that there is a substantial asymmetry in the (nonlocal interactions of the core hole and valence $d$ electrons, based on the comparison of X-ray photoemission spectroscopy and optical conductivity data \cite{okada1997intersite}.  The opposite limit would be  $U_{cp}=U_{dp}$, such that electrons on the $p$ orbitals interact with the charge on the entire copper atom (including core orbitals). Since the dominant XAS transition does not change the total charge on the copper atom, in contrast to electron addition spectroscopy, a nonzero $U_{cp}$ would give an additional asymmetry in the photo-induced shifts of the XAS lines and the electronic spectrum. For a full treatment of the case  $U_{cp}\neq 0$ we would need to incorporate the $c$ level into the GW+EDMFT simulation, and also consider screened interactions $W_{cd}$.

In addition to the shift of the main exciton line, we observe additional spectral weight in the XAS spectrum (see in particular the curve for the large photo-doping $\Delta n_d=9\%$). Such changes can generally originate from two effects: First, photo-excitation opens new absorption channels, where the initial state corresponds to an electronically excited state. Analyzing these states in the $d$-$p$ cluster model \eqref{cluster_}, one finds that, depending on the interaction $U_{cd}$, many of the new transitions are nearly degenerate with the excitonic peak, see vertical lines in Fig.~\ref{Fig:xas_U6}(a).  Furthermore, we could again  expect boson satellites due to the dynamical screening. Similar to the discussion in the equilibrium case (Sec.~\ref{xasequi}), one can try to estimate such boson satellites from the minimal model: After the photo-excitation, additional spectral weight appears in  Im$\,W_{dd}$ around $\omega\approx \Omega_{0}=1$ eV (see Fig.~\ref{Fig:Spectral_W}). If, for a rough estimate, this weight is represented by an oscillator at $\Omega_1=1$~eV with a coupling  $2\gamma_1^2/\Omega_{1} = \Delta W_{dd}(0)$, the system is still in the anti-adiabatic regime (e.g., $\Delta W_{dd}(0)\approx -0.1$~eV for $\Delta n_d=4\%$,  such that $\gamma_1^2/\Omega_{1}^2=0.05$). 
Hence, for the photo-doped charge transfer insulator, we expect that these satellite features are still weak, such that at most the first side-peak may be visible. The latter, however, falls within the linewidth of the exciton. Both the additional XAS channels and the weak boson side-peaks will thus add up to a change of the lineshape of the main exciton. Indeed, the difference plot in Fig.~\ref{Fig:xas_U6}(b) shows that the exciton is not only shifted. Due
to the large intrinsic exciton linewidth, we cannot resolve the nature of the changes in the lineshape within the given accuracy. A possible perspective is to use  resonant inelastic X-ray scattering~(RIXS) to analyze charge fluctuations as these provide direct insights into excited bosonic spectrum.

\section{Conclusions}
\label{Sec:conclusion}

In conclusion, we have addressed how photoinduced changes in charge transfer insulators are encoded in the  time-resolved XAS signal. In particular, after photodoping of charge-transfer insulators, bandgap renormalization is a well-established phenomenon observed in optical~\cite{,matsuda1994,okamoto2007,okamoto2011,novelli12} and photoemission~\cite{cilento2018} spectroscopy. Our work shows that XAS is also a sensitive measure of the dynamical renormalization of the Hubbard $U$,
in line with the interpretation of recent experiments \cite{baykusheva2022}, and we analyzed in some detail how the screening is encoded in the time-resolved XAS signal. 

One important conclusion is that different experimental probes of the interaction renormalization, like  a measurement of the upper Hubbard band using  inverse photoemission~\cite{smith1988} or two-photon photoemission~\cite{gillmeister2020,petek1997}, can respond differently to photo-induced changes in screening as compared to the excitonic XAS transitions. This should be contrasted to the response in the atomic limit, where both depend on $U$ in the same way. Our study therefore suggests that comparison between different spectroscopic probes for the dynamical screening in CT insulators will be very helpful for identifying the nature of photo-induced screening. Further, one should keep in mind that in multiorbital systems there are several mechanisms contributing to the band gap renormalization, in particular the  retarded dynamical screening and the instantaneous Hartree shifts, which then depends on several microscopic parameters.

In principle, dynamical screening is also reflected both in a change of the position and the lineshape of the XAS exciton. We exemplified this using a minimal, exactly solvable model. In the adiabatic limit, 
the  width of the XAS exciton  does not only depend on the core hole lifetime, but also on the coupling and the frequency of the bosonic modes.
In the anti-adiabatic limit, the main effects are excitonic redshifts and the build-up of side peaks at multiples of the bosonic energies. Compared with the full GW+EMDFT simulation, we find that for the CT insulator, the changes in the excitonic lineshapes are weak, because the screening modes have a frequency larger than the induced shifts (anti-adiabatic regime). Besides the screening modes, the XAS signal can also be modified through photo-activated transitions between cluster many-body states, leading to a shoulder-like structure at the excitonic edge and an enhancement of the continuum. 

While the signal of the dynamical screening and photo-activated transitions in XAS is rather weak, RIXS might be a better tool to observe the boson sidepeaks. An extension of the formalism to RIXS~\cite{ament2011,chen2019,eckstein2021,werner2021} would enable a direct analysis of photoinduced changes in screening modes.  Previous studies of electron-lattice coupled systems with a single lattice mode~\cite{werner2021b} showed that one can separate the RIXS signal in photo-doped Mott insulators into the elastic electronic contribution and sidebands originating from the loss of energy to the lattice. A similar separation for the continuum of screening modes in charge transfer insulators would simplify the analysis and provide more direct insights into the origin of dynamical modification of Hubbard $U$ in realistic materials.

Our work also calls for improved formalisms for the treatment of screening and the calculation of the XAS signal. The GW+EDMFT formalism used in the present work can lead to a noncausal action, which could be avoided by a nonequilibrium extension of the  modified GW+EDMFT formalism proposed in Refs.~\onlinecite{backes2022,chen2022}. A further theoretical problem is 
the solution of the time-dependent impurity model, for which we employed the NCA approximation. 
A comparison with an exact solution of the atomic problem, see App.~\ref{app:nca_harm-screen}, reveals substantial deviations if the system is away from the adiabatic or anti-adiabatic limit. These issues should be overcome by recent improvements in steady-state impurity solvers~\cite{erpenbeck2023,kunzel2024} which need to be extended to include retarded density-density interactions.

\acknowledgements
We acknowledge discussions with Matteo Mitrano,  and with Andrea Eschenlohr and Uwe Bovensiepen within the Project P6 of the research group QUAST-FOR5249 - 449872909.
D. G. is supported by the Slovenian Research Agency (ARRS) under Program No. P1-0044, No. J1-2455 and  MN-0016-106. P.~W. acknowledges support from SNSF Grant No. 200021-196966. M.~E. and E.~P. are supported by the Cluster of Excellence „CUI: Advanced Imaging of Matter“ of the Deutsche Forschungsgemeinschaft (DFG) – EXC 2056 – project ID 390715994. 

\appendix
\section{XAS for the minimal model}
\label{app:derivation_xas_minimal-model}

\begin{figure*}[ht]
	\includegraphics[width=.8\linewidth]{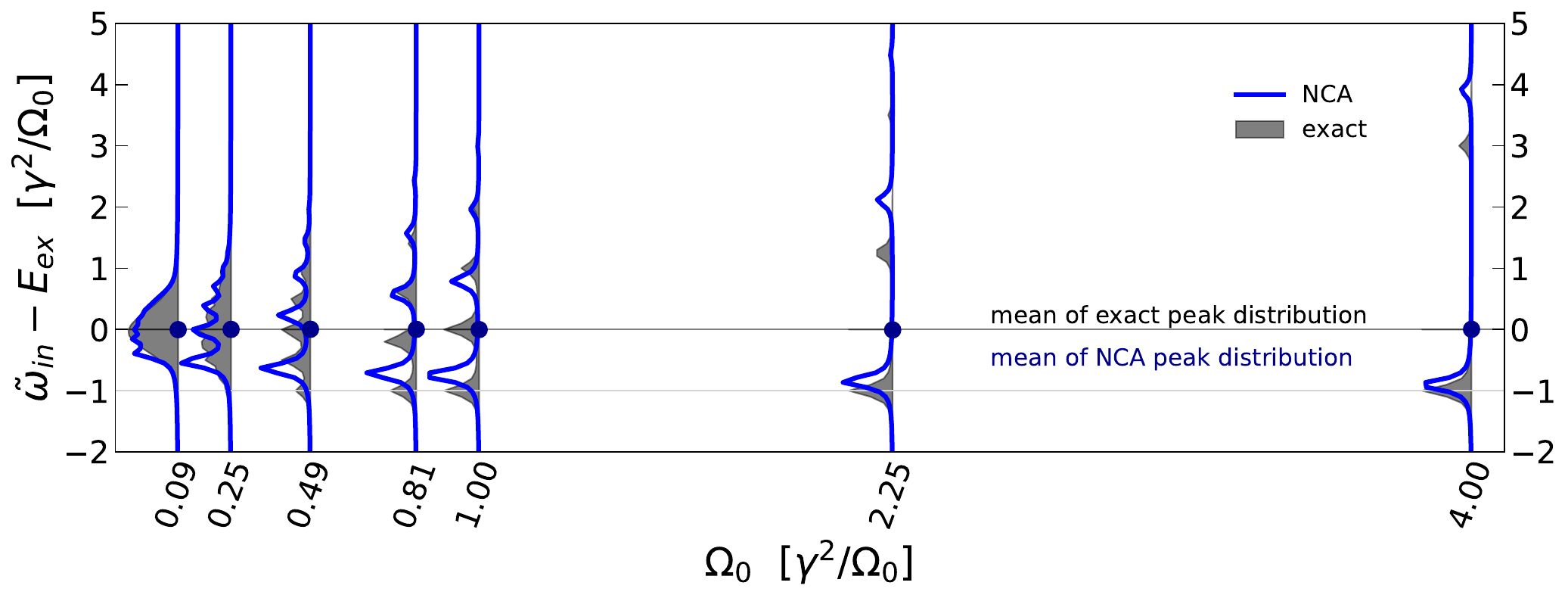}\\
	\caption{XAS signals for a single $d$-orbital, coupled to a harmonic screening mode according to Eqs.~\eqref{eq:xas_harm-screen} and \eqref{eq:int_harm-screen} in the exact ($T=0$) and NCA ($T=1/(50 k_B)$) description, respectively, for various screening frequencies $\Omega_0$. The unit of energy is set to $\gamma^2/\Omega_0$. The light gray line at $-\gamma^2/\Omega_0$ gives the static redshift induced by the presence of the harmonic screening mode. The dark gray line (blue dots) marks the mean of the XAS peak distribution according to the exact (NCA) calculation. The inverse core-hole lifetime is $\Gamma=0.1$. 
	} \label{fig:nca_vs_exact_harm-screen}
	\end{figure*}%

In the following, we sketch the derivation of the XAS formula \eqref{eq:xas_harm-screen} for the minimal model \eqref{eq:H_harm-screen}.  First, we set the dipole matrix elements to $D_{c\sigma,d\sigma'}\equiv \delta_{\sigma,\sigma'}$, such that  the dipolar transition operator simplifies to $P^\dagger = \sum_\sigma d^\dagger_\sigma c_\sigma$. Second, for the minimal model, where the initial state $\ket{\psi_i}$ is an eigenstate (with energy $E_i=2\epsilon_c+\epsilon_d$) of the Hamiltonian $H$, we can modify the expression \eqref{eq:I_XAS_IMP1} for the XAS signal to yield
\begin{align}\begin{split}
	I_{\text{XAS}} =& \int dt_1 \int dt_2 \, s(t_1)s(t_2)\, e^{i(\omega_{\text{in}}+E_i)(t_2-t_1)} \\
	& \quad \times \langle \psi_i | P\, e^{-iH(t_2-t_1)} e^{-\Gamma|t_2-t_1|} P^\dagger  | \psi_i\rangle
	,
\end{split}
	\label{eq:I_XAS_conservative}
\end{align}
assuming a real-valued probe pulse envelope $s(t)$. This is a well-known expression for the XAS signal in a wavefunction-based description~\cite{chen2019}.
Furthermore, since the Hamiltonian \eqref{eq:H_harm-screen} of the minimal model is time-independent, we will recover the equilibrium version of the XAS formula.

From now on we set $s(t)\equiv 1$, assuming an infinite X-ray probe window. Then, formally, expression \eqref{eq:I_XAS_conservative} will diverge, however, the XAS rate, i.e.~the measured number of photons per unit time, is finite. We will focus on the XAS rate for which only one integral over the time difference $\Delta t=t_2-t_1$ in expression \eqref{eq:I_XAS_conservative} remains.
In the final state after the X-ray absorption, additional screening modes will be excited due to the coupling $\propto\gamma\overline{n}_d X$. We calculate the energy of the final state by invoking a Lang-Firsov transform~\cite{lang1962}, which is a unitary transform with (anti-hermitian) generator $S=\frac{\gamma}{\Omega_0}\overline n_d(b-b^\dagger)$. It transforms the Hamiltonian to $\tilde H = e^{S}He^{-S} = \epsilon_d n_d+\epsilon_c n_c +Un_{d\uparrow}n_{d\downarrow} +\Omega_0 b^\dagger b-\frac{\gamma^2}{\Omega_0}\overline n_d^2$, thus removing the explicit electron-boson coupling.
With this, the XAS rate becomes
\begin{align}
	I_{\text{XAS}} =& \int d\Delta t \, e^{i(\omega_{\text{in}}+E_i)\Delta t} e^{-\Gamma|\Delta t|} \nonumber \\
	&\quad \times \langle \psi_i | P e^{-S} e^{S} e^{-iH\Delta t} e^{-S} e^{S}  P^\dagger  | \psi_i\rangle \nonumber \\
	 =& \int d\Delta t\, e^{i(\omega_{\text{in}}+E_i)\Delta t-\Gamma |\Delta t|}\langle \psi_i | P e^{-S}e^{-i\tilde H\Delta t} e^{S}  P^\dagger  | \psi_i\rangle.
	\label{eq:I_XAS_der_minmodel1}
\end{align}
Expanding the X-ray excited state into energy eigenstates $\ket{\tilde\psi_f}$ of $\tilde H$ and performing the integral over $\Delta t$, one finds
\begin{align}
	I_{\text{XAS}} = \sum_f \frac{2\Gamma}{(\omega_{\text{in}}+E_i-\tilde E_f)^2 +\Gamma^2} \Big| \langle \psi_i | P e^{-S} | \tilde\psi_f \rangle \Big|^2 .
	\label{eq:I_XAS_der_minmodel2}
\end{align}
The electronic occupation of the initial and final states $\ket{\psi_i},\ket{\tilde \psi_f}$ is fixed, hence the remaining degree of freedom to sum over in the final state is the screening mode excitation, i.e.~the occupation of the bosonic state. If $\ket{m}_{\text{bos}} = \frac{\left(b^\dagger\right)^m }{\sqrt{m!}} \ket{0}_{\text{bos}}$ is the bosonic state with $m$ screening mode excitations, then the scalar product in \eqref{eq:I_XAS_der_minmodel2} can be written as $\langle0|e^{-(\gamma/\Omega_0)(b-b^\dagger)}|m \rangle_{\text{bos}} = (-\gamma/\Omega_0)^m e^{-\gamma^2/2\Omega_0^2} /\sqrt{m!}$, which is the projection of a coherent state onto the $m$th occupation number state. Inserting this into the expression for $I_{\text{XAS}}$ above yields
\begin{align}
	I_{\text{XAS}} = \sum_{m=0}^\infty \frac{2\Gamma \cdot e^{-\gamma^2/\Omega_0^2} (\gamma/\Omega_0)^{2m} / m! }{(\omega_{\text{in}}+\epsilon_c-\epsilon_d-U-\Omega_0 m +\gamma^2/\Omega_0)^2 +\Gamma^2},
	\label{eq:I_XAS_der_minmodel3}
\end{align}
which is the XAS formula \eqref{eq:xas_harm-screen} in the main text.

\section{Harmonic screening within the non-crossing approximation}
\label{app:nca_harm-screen}

In this appendix we benchmark the non-crossing approximation (NCA) for a harmonic screening mode on the XAS spectrum. We show that in the adiabatic ($\Omega_0\ll\gamma$) and anti-adiabatic ($\Omega_0\gg\gamma$) limit  the NCA reproduces the changes in the XAS signal (in terms of sidebands and excitonic peak shift) reasonably well.

To do so, we set up an impurity formulation of the harmonic screening model \eqref{eq:H_harm-screen} which is obtained by integrating out the bosonic degrees of freedom, generating an effective interaction \eqref{Uret} for the electronic density fluctuations on the $d$-orbital. As discussed in Sec.~\ref{Sec:XAS_har}, the impurity action then takes the same form as in the full  GW+EDMFT  description, see Eqs.~\eqref{actionfuxes2},\eqref{SGWD} and \eqref{envU}, with vanishing hybridization and Hartree self energy on the $d$-orbital, $\Delta(t,t')=0$ and $S_{\mathrm{mf}}=0$. The local Hamiltonian for $S_{\text{loc}}$ is $H_{\mathrm{loc}}= \epsilon_d n_d + \epsilon_c n_c+U n_{d,\uparrow}n_{d,\downarrow}$ and the effective interaction $\mathcal D(t,t')$ is not selfconsistent but derived from the interaction with the screening mode, 
\begin{align}
	\mathcal D(t,t') = -2i \gamma^2 \Omega_0 \langle T_{\mathcal C} X(t) X(t') \rangle,
	\label{eq:int_harm-screen}
\end{align}
c.f.~Eq.~\eqref{Uret}.
The NCA marks the leading order self-consistent expansion of physical observables in in $\mathcal{D}$, see Ref.~\onlinecite{golez2015}. 
Otherwise, the XAS signal is evaluated as described in Sec.~\ref{ssec:XASimp}, in particular Eq.~\eqref{eq:I_XAS_IMP1}, with constant dipole matrix elements $D_{c\sigma,d\sigma'}\equiv \delta_{\sigma,\sigma'}$ and a constant probe pulse envelope $s(t)\equiv 1$. A finite probe pulse envelope will further coarsen the spectral resolution of XAS, which we therefore ignore in this discussion in order to focus on the screening effects. One can further simplify the calculation by taking into account that, for the specific initial state with a singly occupied $d$-orbital, only the greater Keldysh component of $\mathcal D$ contributes in the strong-coupling hybridization expansion.

In Fig.~\ref{fig:nca_vs_exact_harm-screen}, we show the XAS signal for different values of the screening frequency $\Omega_0$, while fixing the static redshift $\gamma^2/\Omega_0 \equiv 1$, as obtained in the exact solution. The NCA results reproduce the broadening effect in the adiabatic ($\Omega_0 \ll \gamma$) and the effective redshift in the anti-adiabtic ($\Omega_0 \gg \gamma$) limit well. The mean of both peak distributions agrees for the entire parameter range.
On the other hand, in the intermediate regime ($\Omega_0 \sim \gamma$), neither the static redshift nor the position of the screening sidebands are reproduced correctly with the NCA.
This suggests that in the adiabatic and anti-adiabatic regimes, we can extend the interpretation of screening effects on XAS in the minimal model (Sec.~\ref{Sec:XAS_har}) to the GW+EDMFT calculation presented in Section \ref{sec:results} of this article.

\section{Cluster analysis} \label{app:cluster}

In this appendix, we provide further details on the atomic limit analysis. The solution of the whole cluster is not exactly solvable for any number 
of holes and therefore we will consider two cases where we can  explicitely analyze all possible XAS transitions: a) the atomic limit with $t_{dd}=t_{dp}=t_{pp}=0$;  b) a mapping of the CuO$_{4}$ cluster to a two-band model, which allows to analyze the effect of nonzero $t_{dp}$ and $t_{pp}$ by considering only the bonding combination of copper and oxygen orbitals.

First, we describe how to map the CuO$_{4}$ cluster  to the two-band problem. We will retain only the symmetrized~(B$_{1g}$) version of the oxygen orbitals, $p_{b}=(c_{p_{x+}}-c_{p_{y+}}-c_{p_{x-}}+c_{p_{y-}})/2$, as these hybridize with the copper $d_{x^2-y^2}$ orbital and neglect other symmetry sectors~\cite{zhang1988,zannen1988,ramsak1989,feiner1996,Jefferson1992}. This leads to the Hamiltonian
\begin{align}
	\begin{split}
  H_0=&\epsilon_d n_{d} +\tilde \epsilon_p n_{p_b} + U_{dd} n_{d,\downarrow} n_{d,\uparrow}
  - \tilde t_{dp} \sum_{\sigma} (c_{d,\sigma}^{\dagger} c_{p_b,\sigma}+h.c.)  \\
  &+U_{dp}/2 n_{d} n_{p_b}+\tilde \epsilon_c n_c + U_{cd} n_d (n_c-2),
	\end{split}
\end{align}
where $\tilde \epsilon_p=\epsilon_p-1.5 t_{pp}$ and $\tilde t_{dp}=2 t_{dp}.$

\paragraph{Transition 1$_d$2$_p$ $\rightarrow$ \underline{2$_d$}2$_p$.}

In the atomic limit~($t_{dp}=t_{p}=0$), the energy of the transitions is given by $\omega=E_\text{tot,at}(\underline{2_d}2_p,1)-E_\text{tot,at}(1_d2_p,2)=U_{dd}/2-\epsilon_c -2 U_{cd}.$ After the inclusion of the intercluster hopping $t_{dp}$, the transition is shifted to $\omega=E_{4,0,0}-E_{3,1/2,1/2}\approx U_{dd}/2-\epsilon_c -2 U_{cd} +\left[4 \frac{t_{dp}^2}{|U_{dd}/2-U_{dp}-\epsilon_p|}\right],$ where the square brackets emphasize the difference with respect to the atomic limit. 

\paragraph{Transition  1$_d$1$_p$ $\rightarrow$ \underline{2$_d$}1$_p$.}
Similarly, the transition from  1$_d$1$_p$ $\rightarrow \underline{2_d}1_p$ is in the atomic limit given by $\omega=E_\text{tot,at}(\underline{2_d}1_p,1)-E_\text{tot,at}(1_d1_p,2)=U_{dd}/2-\epsilon_c -2 U_{cd}-U_{dp}/2.$ After the inclusion of the $t_{dp}$ hybridization, the transition is shifted to $\omega_T=E_{3,1/2,1/2}-E_{2,1,1}=E_{3,1/2,1/2}-E_{2,0,1}\approx U_{dd}/2-\epsilon_c-2U_{cd}-U_{dp}/2+\left[4\frac{t_{dp}^2}{|U_{dd}/2-U_{dp}-\epsilon_p|}\right]$. The main difference with respect to the previous  transition is the shift by $-U_{dp}/2$ due to different occupation on the oxygen orbital. In the singlet channel, the transition is given by  $\omega_S=E_{3,1/2,1/2}-E_{2,0,0}\approx U_{dd}/2-\epsilon_c-2U_{cd}-U_{dp}/2  + \left[ 4\frac{t_{dp}^2}{|U_{dd}/2-U_{dp}-\epsilon_p|}+8\frac{t_{dp}^2}{|U_{dd}/2+\epsilon_p+U_{dp}|} -8\frac{t_{dp}^2}{|3U_{dd}/2+3U_{dp}/2+\epsilon_p|} \right]$.

\paragraph{0$_d$2$_p$ $\rightarrow$ \underline{1$_d$} 2$_p$.}
In the atomic limit of the main text, the energy of the transition is given by $\omega=E_\text{tot,at}(1_d2_p,1)-E_\text{tot,at}(0_d2_p,2)=-U_{dd}/2-\epsilon_c-U_{cd}$. In the cluster model, the transition is renormalized both due to the finite $d$-$p$ interaction and virtual hopping leads to the renormalized frequency $\omega=E_{3,1/2,1/2}-E_{2,0,0}=-U_{dd}/2-\epsilon_c-U_{cd}-4 t_{dp}^2 [\frac{1}{U_{dd}/2-U_{dp}-\epsilon_p}+\frac{2}{U_{dd}/2+3U_{dp}/2+\epsilon_p}].$

\paragraph{0d 1p $\rightarrow$ \underline{0d} 2p.}
The atomic limit transition line is given by $\omega=E_\text{tot,at}(0_d2_p,1)-E_\text{tot,at}(0_d1_p,1)=\epsilon_p-\epsilon_c$. This line gets renormalized by the treatment of the whole cluster as $\omega=E_{2,0,0}-E_{1,1/2,1/2}=\epsilon_p-\epsilon_c-4\frac{t_{dp}^2}{U_{dd}/2+U_{dp}+\epsilon_p}+8\frac{t_{dp}^2}{U_{dd}/2+3U_{dp}/2+\epsilon_p}.$

\paragraph{1d 1p $\rightarrow$ \underline{1d} 2p.}
In the atomic limit the transition is given by $\omega=E_\text{tot,at}(1_d2_p)-E_\text{tot,at}(1_d1_p)=\epsilon_p-\epsilon_c-U_{cd}+U_{dp}/2.$ Within the cluster two transition from the triplet state are degenerate $\omega_T=E_{3,1/2,1/2}-E_{2,1,1}=E_{3,1/2,1/2}-E_{2,0,1}=-\epsilon_c-U_{cd}+U_{dp}/2-4 \frac{t_{dp}^2}{U_{dd}/2-U_{dp}-\epsilon_p}.$ The transition from the singlet state is $\omega_S=E_{3,1/2,1/2}-E_{2,0,0}=-\epsilon_c-U_{cd}+U_{dp}/2-4\frac{t_{dp}^2}{U_{dd}/2-U_{dp}-\epsilon_p}+8\frac{t_{dp}^2}{U_{dd}/2+U_{dp}+\epsilon_p}-8\frac{t_{dp}^2}{3U_{dd}/2+3U_{dp}/2+\epsilon_p}.$ As expected the induced splitting between the singlet and the triplet state is given by different second-order processes between the $d$ and $p$ orbital which for the given parameters are very small numbers and not accessible within the current numerical analysis.

\paragraph{0$_d$ 2$_p$ $\rightarrow$ \underline{2$_d$} 1$_p$.}
The transition line in the atomic limit is given by $\omega=E_\text{tot,at}(2_d 1_p,2)-E_\text{tot,at}(0_d 2_p ,1)=-\epsilon_p-U_{dp}-2U_{cd}-\epsilon_c$. 
The treatment of the whole $d$-$p$ cluster leads  to the renormalization of the resonance as $\omega=E_{3,1/2,1/2}-E_{2,0,0}=-\epsilon_p-U_{dp}-2U_{cd}-\epsilon_c+4\frac{t_{dp}^2}{U_{dd}/2-U_{dp}-\epsilon_p}-8\frac{t_{dp}^2}{U_{dd}/2+3U_{dp}/2+\epsilon_p}.$

\paragraph{2d 1p $\rightarrow$ \underline{2d} 2p.}
The transition line in the atomic limit is given by $\omega=E_\text{tot,at}(2_d2_p)-E_\text{tot,at}(2_d1_p)=-\epsilon_c+\epsilon_p+U_{dp}-2U_{cd}$. The treatment of the whole $d$-$p$ cluster leads to the renormalization of the resonance as $\omega=E_{4,0,0}-E_{3,1/2,1/2}=-\epsilon_c+\epsilon_p+U_{dp}-2U_{cd}-4\frac{t_{dp}^2}{U_{dd}/2-U_{dp}-\epsilon_p}$.

\begin{table*}
\begin{tabular}{c|c|p{8cm}}
\hline
$\alpha~(N,S_z,S))$ & $|j\rangle\in\alpha$ & $E_{\alpha}$\\
\hline\hline
$(0,0,0)$ \hspace{5mm} & $|0,0\rangle$ & 0\\
\hline
$(1,1/2,1/2)$ & $\{|\uparrow,0\rangle$, $|0,\uparrow\rangle$\} & $\frac{(-U_{dd}/2-U_{dp}+\epsilon_p)}{2}\pm\sqrt{\left(\frac{U_{dd}/2+U_{dp}+\epsilon_p}{2}\right)^2+4t_{dp}^2}$\\
\hline
$(2,1,1)$ & $|\uparrow,\uparrow\rangle$ &  $-U_{dd}/2+\epsilon_p-U_{dp}/2$\\
\hline
$(2,0,0)$ & $\{|\!\uparrow\downarrow,0\rangle$,$|0,\uparrow\downarrow\rangle$ $,|\!\downarrow,\uparrow\rangle-|\!\uparrow,\downarrow\rangle$\} & $ -\frac{8 t_{dp}^2}{-U_{dd}/2-3U_{dp}/2-\epsilon_p}+2\epsilon_p$ \newline $-U_{dd}/2-U_{dp}/2+\epsilon_p -\frac{8 t_{dp}^2}{\epsilon_p+U_{dd}+U_{dp}}+\frac{8 t_{dp}^2}{\epsilon_p+3U_{dd}/2+3U_{dp}/2}$ \newline 
$-2 U_{dp}-\frac{8 t_{dp}^2}{\epsilon_p+3(U_{dd}+U_{dp})/2}$ \\
\hline
$(2,0,1)$ & $\{|\!\downarrow,\uparrow\rangle+|\!\uparrow,\downarrow\rangle$\} & $-U_{dd}/2+\epsilon_p-U_{dp}/2$\\
\hline
$(3,1/2,1/2)$ & $\{|\!\uparrow,\uparrow\downarrow\rangle$, $|\!\uparrow\downarrow,\uparrow\rangle$\} & $\frac{(-U_{dd}/2-U_{dp}+3\epsilon_p)}{2} \pm \sqrt{\left (\frac{U_{dd}/2-U_{dp}-\epsilon_p}{2}\right )^2+4t_{dp}^2}$ \\
\hline
$(4,0,0)$ & $|\!\uparrow\downarrow,\uparrow\downarrow\rangle$\hspace{5mm} & $2 \epsilon_p$\\
\hline
\end{tabular}
\caption{Labeling of states in the $d$-$p$ cluster, where states are sorted by the total number of electrons $N$, $z$-component of the total spin $S_z$ and the total spin $S$. For states with a nonzero value of $S_z$ we show only fully polarized states. The second column shows the $d$-$p$ states belonging to a given multiplet $\alpha$, where the first (second) entry in the ket refers to the hole occupation of orbital $d$ ($p_+$). The last column shows the energy of the states except for (N,S$_z$,S)=(2,0,0), where the analytical expressions are too complicated, and energies are calculated within second order perturbation theory.}
\label{tabstates02}
\end{table*}

\begin{thebibliography}{63}%
\makeatletter
\providecommand \@ifxundefined [1]{%
 \@ifx{#1\undefined}
}%
\providecommand \@ifnum [1]{%
 \ifnum #1\expandafter \@firstoftwo
 \else \expandafter \@secondoftwo
 \fi
}%
\providecommand \@ifx [1]{%
 \ifx #1\expandafter \@firstoftwo
 \else \expandafter \@secondoftwo
 \fi
}%
\providecommand \natexlab [1]{#1}%
\providecommand \enquote  [1]{``#1''}%
\providecommand \bibnamefont  [1]{#1}%
\providecommand \bibfnamefont [1]{#1}%
\providecommand \citenamefont [1]{#1}%
\providecommand \href@noop [0]{\@secondoftwo}%
\providecommand \href [0]{\begingroup \@sanitize@url \@href}%
\providecommand \@href[1]{\@@startlink{#1}\@@href}%
\providecommand \@@href[1]{\endgroup#1\@@endlink}%
\providecommand \@sanitize@url [0]{\catcode `\\12\catcode `\$12\catcode
  `\&12\catcode `\#12\catcode `\^12\catcode `\_12\catcode `\%12\relax}%
\providecommand \@@startlink[1]{}%
\providecommand \@@endlink[0]{}%
\providecommand \url  [0]{\begingroup\@sanitize@url \@url }%
\providecommand \@url [1]{\endgroup\@href {#1}{\urlprefix }}%
\providecommand \urlprefix  [0]{URL }%
\providecommand \Eprint [0]{\href }%
\providecommand \doibase [0]{https://doi.org/}%
\providecommand \selectlanguage [0]{\@gobble}%
\providecommand \bibinfo  [0]{\@secondoftwo}%
\providecommand \bibfield  [0]{\@secondoftwo}%
\providecommand \translation [1]{[#1]}%
\providecommand \BibitemOpen [0]{}%
\providecommand \bibitemStop [0]{}%
\providecommand \bibitemNoStop [0]{.\EOS\space}%
\providecommand \EOS [0]{\spacefactor3000\relax}%
\providecommand \BibitemShut  [1]{\csname bibitem#1\endcsname}%
\let\auto@bib@innerbib\@empty
\bibitem [{\citenamefont {de~Groot}\ \emph {et~al.}(2021)\citenamefont
  {de~Groot}, \citenamefont {Elnaggar}, \citenamefont {Frati}, \citenamefont
  {Wang}, \citenamefont {Delgado-Jaime}, \citenamefont {van Veenendaal},
  \citenamefont {Fernandez-Rodriguez}, \citenamefont {Haverkort}, \citenamefont
  {Green}, \citenamefont {van~der Laan} \emph {et~al.}}]{de20212p}%
  \BibitemOpen
  \bibfield  {author} {\bibinfo {author} {\bibfnamefont {F.~M.}\ \bibnamefont
  {de~Groot}}, \bibinfo {author} {\bibfnamefont {H.}~\bibnamefont {Elnaggar}},
  \bibinfo {author} {\bibfnamefont {F.}~\bibnamefont {Frati}}, \bibinfo
  {author} {\bibfnamefont {R.-p.}\ \bibnamefont {Wang}}, \bibinfo {author}
  {\bibfnamefont {M.~U.}\ \bibnamefont {Delgado-Jaime}}, \bibinfo {author}
  {\bibfnamefont {M.}~\bibnamefont {van Veenendaal}}, \bibinfo {author}
  {\bibfnamefont {J.}~\bibnamefont {Fernandez-Rodriguez}}, \bibinfo {author}
  {\bibfnamefont {M.~W.}\ \bibnamefont {Haverkort}}, \bibinfo {author}
  {\bibfnamefont {R.~J.}\ \bibnamefont {Green}}, \bibinfo {author}
  {\bibfnamefont {G.}~\bibnamefont {van~der Laan}}, \emph {et~al.},\ }\bibfield
   {title} {\bibinfo {title} {2p x-ray absorption spectroscopy of 3d transition
  metal systems},\ }\href@noop {} {\bibfield  {journal} {\bibinfo  {journal}
  {Journal of Electron Spectroscopy and Related Phenomena}\ }\textbf {\bibinfo
  {volume} {249}},\ \bibinfo {pages} {147061} (\bibinfo {year}
  {2021})}\BibitemShut {NoStop}%
\bibitem [{\citenamefont {Ament}\ \emph {et~al.}(2011)\citenamefont {Ament},
  \citenamefont {van Veenendaal}, \citenamefont {Devereaux}, \citenamefont
  {Hill},\ and\ \citenamefont {van~den Brink}}]{ament2011}%
  \BibitemOpen
  \bibfield  {author} {\bibinfo {author} {\bibfnamefont {L.~J.~P.}\
  \bibnamefont {Ament}}, \bibinfo {author} {\bibfnamefont {M.}~\bibnamefont
  {van Veenendaal}}, \bibinfo {author} {\bibfnamefont {T.~P.}\ \bibnamefont
  {Devereaux}}, \bibinfo {author} {\bibfnamefont {J.~P.}\ \bibnamefont
  {Hill}},\ and\ \bibinfo {author} {\bibfnamefont {J.}~\bibnamefont {van~den
  Brink}},\ }\bibfield  {title} {\bibinfo {title} {Resonant inelastic x-ray
  scattering studies of elementary excitations},\ }\href
  {https://doi.org/10.1103/RevModPhys.83.705} {\bibfield  {journal} {\bibinfo
  {journal} {Rev. Mod. Phys.}\ }\textbf {\bibinfo {volume} {83}},\ \bibinfo
  {pages} {705} (\bibinfo {year} {2011})}\BibitemShut {NoStop}%
\bibitem [{\citenamefont {Kuo}\ \emph {et~al.}(2017)\citenamefont {Kuo},
  \citenamefont {Haupricht}, \citenamefont {Weinen}, \citenamefont {Wu},
  \citenamefont {Tsuei}, \citenamefont {Haverkort}, \citenamefont {Tanaka},\
  and\ \citenamefont {Tjeng}}]{kuo2017challenges}%
  \BibitemOpen
  \bibfield  {author} {\bibinfo {author} {\bibfnamefont {C.~Y.}\ \bibnamefont
  {Kuo}}, \bibinfo {author} {\bibfnamefont {T.}~\bibnamefont {Haupricht}},
  \bibinfo {author} {\bibfnamefont {J.}~\bibnamefont {Weinen}}, \bibinfo
  {author} {\bibfnamefont {H.}~\bibnamefont {Wu}}, \bibinfo {author}
  {\bibfnamefont {K.~D.}\ \bibnamefont {Tsuei}}, \bibinfo {author}
  {\bibfnamefont {M.}~\bibnamefont {Haverkort}}, \bibinfo {author}
  {\bibfnamefont {A.}~\bibnamefont {Tanaka}},\ and\ \bibinfo {author}
  {\bibfnamefont {L.}~\bibnamefont {Tjeng}},\ }\bibfield  {title} {\bibinfo
  {title} {Challenges from experiment: electronic structure of nio},\
  }\href@noop {} {\bibfield  {journal} {\bibinfo  {journal} {The European
  Physical Journal Special Topics}\ }\textbf {\bibinfo {volume} {226}},\
  \bibinfo {pages} {2445} (\bibinfo {year} {2017})}\BibitemShut {NoStop}%
\bibitem [{\citenamefont {Giannetti}\ \emph {et~al.}(2016)\citenamefont
  {Giannetti}, \citenamefont {Capone}, \citenamefont {Fausti}, \citenamefont
  {Fabrizio}, \citenamefont {Parmigiani},\ and\ \citenamefont
  {Mihailovic}}]{Giannetti2016}%
  \BibitemOpen
  \bibfield  {author} {\bibinfo {author} {\bibfnamefont {C.}~\bibnamefont
  {Giannetti}}, \bibinfo {author} {\bibfnamefont {M.}~\bibnamefont {Capone}},
  \bibinfo {author} {\bibfnamefont {D.}~\bibnamefont {Fausti}}, \bibinfo
  {author} {\bibfnamefont {M.}~\bibnamefont {Fabrizio}}, \bibinfo {author}
  {\bibfnamefont {F.}~\bibnamefont {Parmigiani}},\ and\ \bibinfo {author}
  {\bibfnamefont {D.}~\bibnamefont {Mihailovic}},\ }\bibfield  {title}
  {\bibinfo {title} {Ultrafast optical spectroscopy of strongly correlated
  materials and high-temperature superconductors: a non-equilibrium approach},\
  }\href {https://doi.org/10.1080/00018732.2016.1194044} {\bibfield  {journal}
  {\bibinfo  {journal} {Adv. Phys.}\ }\textbf {\bibinfo {volume} {65}},\
  \bibinfo {pages} {58} (\bibinfo {year} {2016})}\BibitemShut {NoStop}%
\bibitem [{\citenamefont {Basov}\ \emph {et~al.}(2017)\citenamefont {Basov},
  \citenamefont {Averitt},\ and\ \citenamefont {Hsieh}}]{Basov2017}%
  \BibitemOpen
  \bibfield  {author} {\bibinfo {author} {\bibfnamefont {D.~N.}\ \bibnamefont
  {Basov}}, \bibinfo {author} {\bibfnamefont {R.~D.}\ \bibnamefont {Averitt}},\
  and\ \bibinfo {author} {\bibfnamefont {D.}~\bibnamefont {Hsieh}},\ }\bibfield
   {title} {\bibinfo {title} {Towards properties on demand in quantum
  materials},\ }\href {https://doi.org/10.1038/nmat5017} {\bibfield  {journal}
  {\bibinfo  {journal} {Nature Materials}\ }\textbf {\bibinfo {volume} {16}},\
  \bibinfo {pages} {1077} (\bibinfo {year} {2017})}\BibitemShut {NoStop}%
\bibitem [{\citenamefont {de~la Torre}\ \emph {et~al.}(2021)\citenamefont
  {de~la Torre}, \citenamefont {Kennes}, \citenamefont {Claassen},
  \citenamefont {Gerber}, \citenamefont {McIver},\ and\ \citenamefont
  {Sentef}}]{torre2021}%
  \BibitemOpen
  \bibfield  {author} {\bibinfo {author} {\bibfnamefont {A.}~\bibnamefont
  {de~la Torre}}, \bibinfo {author} {\bibfnamefont {D.~M.}\ \bibnamefont
  {Kennes}}, \bibinfo {author} {\bibfnamefont {M.}~\bibnamefont {Claassen}},
  \bibinfo {author} {\bibfnamefont {S.}~\bibnamefont {Gerber}}, \bibinfo
  {author} {\bibfnamefont {J.~W.}\ \bibnamefont {McIver}},\ and\ \bibinfo
  {author} {\bibfnamefont {M.~A.}\ \bibnamefont {Sentef}},\ }\bibfield  {title}
  {\bibinfo {title} {Colloquium: Nonthermal pathways to ultrafast control in
  quantum materials},\ }\href {https://doi.org/10.1103/RevModPhys.93.041002}
  {\bibfield  {journal} {\bibinfo  {journal} {Rev. Mod. Phys.}\ }\textbf
  {\bibinfo {volume} {93}},\ \bibinfo {pages} {041002} (\bibinfo {year}
  {2021})}\BibitemShut {NoStop}%
\bibitem [{\citenamefont {Murakami}\ \emph {et~al.}(2023)\citenamefont
  {Murakami}, \citenamefont {Gole{\v z}}, \citenamefont {Eckstein},\ and\
  \citenamefont {Werner}}]{Murakami2023}%
  \BibitemOpen
  \bibfield  {author} {\bibinfo {author} {\bibfnamefont {Y.}~\bibnamefont
  {Murakami}}, \bibinfo {author} {\bibfnamefont {D.}~\bibnamefont {Gole{\v
  z}}}, \bibinfo {author} {\bibfnamefont {M.}~\bibnamefont {Eckstein}},\ and\
  \bibinfo {author} {\bibfnamefont {P.}~\bibnamefont {Werner}},\ }\href@noop {}
  {\bibinfo {title} {Photo-induced nonequilibrium states in {Mott} insulators}}
  (\bibinfo {year} {2023}),\ \Eprint {https://arxiv.org/abs/2310.05201}
  {arXiv:2310.05201 [cond-mat.str-el]} \BibitemShut {NoStop}%
\bibitem [{\citenamefont {Boschini}\ \emph {et~al.}(2024)\citenamefont
  {Boschini}, \citenamefont {Zonno},\ and\ \citenamefont
  {Damascelli}}]{boschini2024}%
  \BibitemOpen
  \bibfield  {author} {\bibinfo {author} {\bibfnamefont {F.}~\bibnamefont
  {Boschini}}, \bibinfo {author} {\bibfnamefont {M.}~\bibnamefont {Zonno}},\
  and\ \bibinfo {author} {\bibfnamefont {A.}~\bibnamefont {Damascelli}},\
  }\bibfield  {title} {\bibinfo {title} {Time-resolved arpes studies of quantum
  materials},\ }\href {https://doi.org/10.1103/RevModPhys.96.015003} {\bibfield
   {journal} {\bibinfo  {journal} {Rev. Mod. Phys.}\ }\textbf {\bibinfo
  {volume} {96}},\ \bibinfo {pages} {015003} (\bibinfo {year}
  {2024})}\BibitemShut {NoStop}%
\bibitem [{\citenamefont {Baykusheva}\ \emph {et~al.}(2022)\citenamefont
  {Baykusheva}, \citenamefont {Jang}, \citenamefont {Husain}, \citenamefont
  {Lee}, \citenamefont {TenHuisen}, \citenamefont {Zhou}, \citenamefont {Park},
  \citenamefont {Kim}, \citenamefont {Kim}, \citenamefont {Kim}, \citenamefont
  {Kim}, \citenamefont {Park}, \citenamefont {Abbamonte}, \citenamefont {Kim},
  \citenamefont {Gu}, \citenamefont {Wang},\ and\ \citenamefont
  {Mitrano}}]{baykusheva2022}%
  \BibitemOpen
  \bibfield  {author} {\bibinfo {author} {\bibfnamefont {D.~R.}\ \bibnamefont
  {Baykusheva}}, \bibinfo {author} {\bibfnamefont {H.}~\bibnamefont {Jang}},
  \bibinfo {author} {\bibfnamefont {A.~A.}\ \bibnamefont {Husain}}, \bibinfo
  {author} {\bibfnamefont {S.}~\bibnamefont {Lee}}, \bibinfo {author}
  {\bibfnamefont {S.~F.~R.}\ \bibnamefont {TenHuisen}}, \bibinfo {author}
  {\bibfnamefont {P.}~\bibnamefont {Zhou}}, \bibinfo {author} {\bibfnamefont
  {S.}~\bibnamefont {Park}}, \bibinfo {author} {\bibfnamefont {H.}~\bibnamefont
  {Kim}}, \bibinfo {author} {\bibfnamefont {J.-K.}\ \bibnamefont {Kim}},
  \bibinfo {author} {\bibfnamefont {H.-D.}\ \bibnamefont {Kim}}, \bibinfo
  {author} {\bibfnamefont {M.}~\bibnamefont {Kim}}, \bibinfo {author}
  {\bibfnamefont {S.-Y.}\ \bibnamefont {Park}}, \bibinfo {author}
  {\bibfnamefont {P.}~\bibnamefont {Abbamonte}}, \bibinfo {author}
  {\bibfnamefont {B.~J.}\ \bibnamefont {Kim}}, \bibinfo {author} {\bibfnamefont
  {G.~D.}\ \bibnamefont {Gu}}, \bibinfo {author} {\bibfnamefont
  {Y.}~\bibnamefont {Wang}},\ and\ \bibinfo {author} {\bibfnamefont
  {M.}~\bibnamefont {Mitrano}},\ }\bibfield  {title} {\bibinfo {title}
  {Ultrafast renormalization of the on-site coulomb repulsion in a cuprate
  superconductor},\ }\href {https://doi.org/10.1103/PhysRevX.12.011013}
  {\bibfield  {journal} {\bibinfo  {journal} {Phys. Rev. X}\ }\textbf {\bibinfo
  {volume} {12}},\ \bibinfo {pages} {011013} (\bibinfo {year}
  {2022})}\BibitemShut {NoStop}%
\bibitem [{\citenamefont {Wang}\ \emph {et~al.}(2022)\citenamefont {Wang},
  \citenamefont {Engel}, \citenamefont {Vaskivskyi}, \citenamefont {Turenne},
  \citenamefont {Shokeen}, \citenamefont {Yaroslavtsev}, \citenamefont
  {Gr{\aa}n{\"a}s}, \citenamefont {Knut}, \citenamefont {Schunck},
  \citenamefont {Dziarzhytski} \emph {et~al.}}]{wang2022ultrafast}%
  \BibitemOpen
  \bibfield  {author} {\bibinfo {author} {\bibfnamefont {X.}~\bibnamefont
  {Wang}}, \bibinfo {author} {\bibfnamefont {R.~Y.}\ \bibnamefont {Engel}},
  \bibinfo {author} {\bibfnamefont {I.}~\bibnamefont {Vaskivskyi}}, \bibinfo
  {author} {\bibfnamefont {D.}~\bibnamefont {Turenne}}, \bibinfo {author}
  {\bibfnamefont {V.}~\bibnamefont {Shokeen}}, \bibinfo {author} {\bibfnamefont
  {A.}~\bibnamefont {Yaroslavtsev}}, \bibinfo {author} {\bibfnamefont
  {O.}~\bibnamefont {Gr{\aa}n{\"a}s}}, \bibinfo {author} {\bibfnamefont
  {R.}~\bibnamefont {Knut}}, \bibinfo {author} {\bibfnamefont {J.~O.}\
  \bibnamefont {Schunck}}, \bibinfo {author} {\bibfnamefont {S.}~\bibnamefont
  {Dziarzhytski}}, \emph {et~al.},\ }\bibfield  {title} {\bibinfo {title}
  {Ultrafast manipulation of the nio antiferromagnetic order via sub-gap
  optical excitation},\ }\href@noop {} {\bibfield  {journal} {\bibinfo
  {journal} {Faraday discussions}\ }\textbf {\bibinfo {volume} {237}},\
  \bibinfo {pages} {300} (\bibinfo {year} {2022})}\BibitemShut {NoStop}%
\bibitem [{\citenamefont {Gr{\aa}n{\"a}s}\ \emph {et~al.}(2022)\citenamefont
  {Gr{\aa}n{\"a}s}, \citenamefont {Vaskivskyi}, \citenamefont {Wang},
  \citenamefont {Thunstr{\"o}m}, \citenamefont {Ghimire}, \citenamefont {Knut},
  \citenamefont {S{\"o}derstr{\"o}m}, \citenamefont {Kjellsson}, \citenamefont
  {Turenne}, \citenamefont {Engel} \emph {et~al.}}]{graanas2022}%
  \BibitemOpen
  \bibfield  {author} {\bibinfo {author} {\bibfnamefont {O.}~\bibnamefont
  {Gr{\aa}n{\"a}s}}, \bibinfo {author} {\bibfnamefont {I.}~\bibnamefont
  {Vaskivskyi}}, \bibinfo {author} {\bibfnamefont {X.}~\bibnamefont {Wang}},
  \bibinfo {author} {\bibfnamefont {P.}~\bibnamefont {Thunstr{\"o}m}}, \bibinfo
  {author} {\bibfnamefont {S.}~\bibnamefont {Ghimire}}, \bibinfo {author}
  {\bibfnamefont {R.}~\bibnamefont {Knut}}, \bibinfo {author} {\bibfnamefont
  {J.}~\bibnamefont {S{\"o}derstr{\"o}m}}, \bibinfo {author} {\bibfnamefont
  {L.}~\bibnamefont {Kjellsson}}, \bibinfo {author} {\bibfnamefont
  {D.}~\bibnamefont {Turenne}}, \bibinfo {author} {\bibfnamefont
  {R.}~\bibnamefont {Engel}}, \emph {et~al.},\ }\bibfield  {title} {\bibinfo
  {title} {Ultrafast modification of the electronic structure of a correlated
  insulator},\ }\href@noop {} {\bibfield  {journal} {\bibinfo  {journal}
  {Physical Review Research}\ }\textbf {\bibinfo {volume} {4}},\ \bibinfo
  {pages} {L032030} (\bibinfo {year} {2022})}\BibitemShut {NoStop}%
\bibitem [{\citenamefont {Huber}\ \emph {et~al.}(2001)\citenamefont {Huber},
  \citenamefont {Tauser}, \citenamefont {Brodschelm}, \citenamefont {Bichler},
  \citenamefont {Abstreiter},\ and\ \citenamefont {Leitenstorfer}}]{huber2001}%
  \BibitemOpen
  \bibfield  {author} {\bibinfo {author} {\bibfnamefont {R.}~\bibnamefont
  {Huber}}, \bibinfo {author} {\bibfnamefont {F.}~\bibnamefont {Tauser}},
  \bibinfo {author} {\bibfnamefont {A.}~\bibnamefont {Brodschelm}}, \bibinfo
  {author} {\bibfnamefont {M.}~\bibnamefont {Bichler}}, \bibinfo {author}
  {\bibfnamefont {G.}~\bibnamefont {Abstreiter}},\ and\ \bibinfo {author}
  {\bibfnamefont {A.}~\bibnamefont {Leitenstorfer}},\ }\bibfield  {title}
  {\bibinfo {title} {How many-particle interactions develop after ultrafast
  excitation of an electron--hole plasma},\ }\href@noop {} {\bibfield
  {journal} {\bibinfo  {journal} {Nature}\ }\textbf {\bibinfo {volume} {414}},\
  \bibinfo {pages} {286} (\bibinfo {year} {2001})}\BibitemShut {NoStop}%
\bibitem [{\citenamefont {Gole\ifmmode~\check{z}\else \v{z}\fi{}}\ \emph
  {et~al.}(2019{\natexlab{a}})\citenamefont {Gole\ifmmode~\check{z}\else
  \v{z}\fi{}}, \citenamefont {Boehnke}, \citenamefont {Eckstein},\ and\
  \citenamefont {Werner}}]{golez2019}%
  \BibitemOpen
  \bibfield  {author} {\bibinfo {author} {\bibfnamefont {D.}~\bibnamefont
  {Gole\ifmmode~\check{z}\else \v{z}\fi{}}}, \bibinfo {author} {\bibfnamefont
  {L.}~\bibnamefont {Boehnke}}, \bibinfo {author} {\bibfnamefont
  {M.}~\bibnamefont {Eckstein}},\ and\ \bibinfo {author} {\bibfnamefont
  {P.}~\bibnamefont {Werner}},\ }\bibfield  {title} {\bibinfo {title} {Dynamics
  of photodoped charge transfer insulators},\ }\href
  {https://doi.org/10.1103/PhysRevB.100.041111} {\bibfield  {journal} {\bibinfo
   {journal} {Phys. Rev. B}\ }\textbf {\bibinfo {volume} {100}},\ \bibinfo
  {pages} {041111} (\bibinfo {year} {2019}{\natexlab{a}})}\BibitemShut
  {NoStop}%
\bibitem [{\citenamefont {Gole\ifmmode~\check{z}\else \v{z}\fi{}}\ \emph
  {et~al.}(2019{\natexlab{b}})\citenamefont {Gole\ifmmode~\check{z}\else
  \v{z}\fi{}}, \citenamefont {Eckstein},\ and\ \citenamefont
  {Werner}}]{golez2019a}%
  \BibitemOpen
  \bibfield  {author} {\bibinfo {author} {\bibfnamefont {D.}~\bibnamefont
  {Gole\ifmmode~\check{z}\else \v{z}\fi{}}}, \bibinfo {author} {\bibfnamefont
  {M.}~\bibnamefont {Eckstein}},\ and\ \bibinfo {author} {\bibfnamefont
  {P.}~\bibnamefont {Werner}},\ }\bibfield  {title} {\bibinfo {title}
  {Multiband nonequilibrium $gw+\text{EDMFT}$ formalism for correlated
  insulators},\ }\href {https://doi.org/10.1103/PhysRevB.100.235117} {\bibfield
   {journal} {\bibinfo  {journal} {Phys. Rev. B}\ }\textbf {\bibinfo {volume}
  {100}},\ \bibinfo {pages} {235117} (\bibinfo {year}
  {2019}{\natexlab{b}})}\BibitemShut {NoStop}%
\bibitem [{\citenamefont {Tancogne-Dejean}\ \emph {et~al.}(2018)\citenamefont
  {Tancogne-Dejean}, \citenamefont {Sentef},\ and\ \citenamefont
  {Rubio}}]{tancogne2018}%
  \BibitemOpen
  \bibfield  {author} {\bibinfo {author} {\bibfnamefont {N.}~\bibnamefont
  {Tancogne-Dejean}}, \bibinfo {author} {\bibfnamefont {M.~A.}\ \bibnamefont
  {Sentef}},\ and\ \bibinfo {author} {\bibfnamefont {A.}~\bibnamefont
  {Rubio}},\ }\bibfield  {title} {\bibinfo {title} {Ultrafast modification of
  hubbard $u$ in a strongly correlated material: Ab initio high-harmonic
  generation in nio},\ }\href {https://doi.org/10.1103/PhysRevLett.121.097402}
  {\bibfield  {journal} {\bibinfo  {journal} {Phys. Rev. Lett.}\ }\textbf
  {\bibinfo {volume} {121}},\ \bibinfo {pages} {097402} (\bibinfo {year}
  {2018})}\BibitemShut {NoStop}%
\bibitem [{\citenamefont {Lojewski}\ \emph {et~al.}(2024)\citenamefont
  {Lojewski}, \citenamefont {Golez}, \citenamefont {Ollefs}, \citenamefont
  {Guyader}, \citenamefont {Kämmerer}, \citenamefont {Rothenbach},
  \citenamefont {Engel}, \citenamefont {Miedema}, \citenamefont {Beye},
  \citenamefont {Chiuzbăian}, \citenamefont {Carley}, \citenamefont {Gort},
  \citenamefont {Kuiken}, \citenamefont {Mercurio}, \citenamefont {Schlappa},
  \citenamefont {Yaroslavtsev}, \citenamefont {Scherz}, \citenamefont
  {Döring}, \citenamefont {David}, \citenamefont {Wende}, \citenamefont
  {Bovensiepen}, \citenamefont {Eckstein}, \citenamefont {Werner},\ and\
  \citenamefont {Eschenlohr}}]{lojewski2024}%
  \BibitemOpen
  \bibfield  {author} {\bibinfo {author} {\bibfnamefont {T.}~\bibnamefont
  {Lojewski}}, \bibinfo {author} {\bibfnamefont {D.}~\bibnamefont {Golez}},
  \bibinfo {author} {\bibfnamefont {K.}~\bibnamefont {Ollefs}}, \bibinfo
  {author} {\bibfnamefont {L.~L.}\ \bibnamefont {Guyader}}, \bibinfo {author}
  {\bibfnamefont {L.}~\bibnamefont {Kämmerer}}, \bibinfo {author}
  {\bibfnamefont {N.}~\bibnamefont {Rothenbach}}, \bibinfo {author}
  {\bibfnamefont {R.~Y.}\ \bibnamefont {Engel}}, \bibinfo {author}
  {\bibfnamefont {P.~S.}\ \bibnamefont {Miedema}}, \bibinfo {author}
  {\bibfnamefont {M.}~\bibnamefont {Beye}}, \bibinfo {author} {\bibfnamefont
  {G.~S.}\ \bibnamefont {Chiuzbăian}}, \bibinfo {author} {\bibfnamefont
  {R.}~\bibnamefont {Carley}}, \bibinfo {author} {\bibfnamefont
  {R.}~\bibnamefont {Gort}}, \bibinfo {author} {\bibfnamefont {B.~E.~V.}\
  \bibnamefont {Kuiken}}, \bibinfo {author} {\bibfnamefont {G.}~\bibnamefont
  {Mercurio}}, \bibinfo {author} {\bibfnamefont {J.}~\bibnamefont {Schlappa}},
  \bibinfo {author} {\bibfnamefont {A.}~\bibnamefont {Yaroslavtsev}}, \bibinfo
  {author} {\bibfnamefont {A.}~\bibnamefont {Scherz}}, \bibinfo {author}
  {\bibfnamefont {F.}~\bibnamefont {Döring}}, \bibinfo {author} {\bibfnamefont
  {C.}~\bibnamefont {David}}, \bibinfo {author} {\bibfnamefont
  {H.}~\bibnamefont {Wende}}, \bibinfo {author} {\bibfnamefont
  {U.}~\bibnamefont {Bovensiepen}}, \bibinfo {author} {\bibfnamefont
  {M.}~\bibnamefont {Eckstein}}, \bibinfo {author} {\bibfnamefont
  {P.}~\bibnamefont {Werner}},\ and\ \bibinfo {author} {\bibfnamefont
  {A.}~\bibnamefont {Eschenlohr}},\ }\href {https://arxiv.org/abs/2305.10145}
  {\bibinfo {title} {Photo-induced charge-transfer renormalization in nio}}
  (\bibinfo {year} {2024}),\ \Eprint {https://arxiv.org/abs/2305.10145}
  {arXiv:2305.10145 [cond-mat.str-el]} \BibitemShut {NoStop}%
\bibitem [{\citenamefont {Chen}\ \emph {et~al.}(2019)\citenamefont {Chen},
  \citenamefont {Wang}, \citenamefont {Jia}, \citenamefont {Moritz},
  \citenamefont {Shvaika}, \citenamefont {Freericks},\ and\ \citenamefont
  {Devereaux}}]{chen2019}%
  \BibitemOpen
  \bibfield  {author} {\bibinfo {author} {\bibfnamefont {Y.}~\bibnamefont
  {Chen}}, \bibinfo {author} {\bibfnamefont {Y.}~\bibnamefont {Wang}}, \bibinfo
  {author} {\bibfnamefont {C.}~\bibnamefont {Jia}}, \bibinfo {author}
  {\bibfnamefont {B.}~\bibnamefont {Moritz}}, \bibinfo {author} {\bibfnamefont
  {A.~M.}\ \bibnamefont {Shvaika}}, \bibinfo {author} {\bibfnamefont {J.~K.}\
  \bibnamefont {Freericks}},\ and\ \bibinfo {author} {\bibfnamefont {T.~P.}\
  \bibnamefont {Devereaux}},\ }\bibfield  {title} {\bibinfo {title} {Theory for
  time-resolved resonant inelastic x-ray scattering},\ }\href
  {https://doi.org/10.1103/PhysRevB.99.104306} {\bibfield  {journal} {\bibinfo
  {journal} {Phys. Rev. B}\ }\textbf {\bibinfo {volume} {99}},\ \bibinfo
  {pages} {104306} (\bibinfo {year} {2019})}\BibitemShut {NoStop}%
\bibitem [{\citenamefont {Werner}\ \emph {et~al.}(2022)\citenamefont {Werner},
  \citenamefont {Golez},\ and\ \citenamefont {Eckstein}}]{werner2022}%
  \BibitemOpen
  \bibfield  {author} {\bibinfo {author} {\bibfnamefont {P.}~\bibnamefont
  {Werner}}, \bibinfo {author} {\bibfnamefont {D.}~\bibnamefont {Golez}},\ and\
  \bibinfo {author} {\bibfnamefont {M.}~\bibnamefont {Eckstein}},\ }\bibfield
  {title} {\bibinfo {title} {Local interpretation of time-resolved x-ray
  absorption in mott insulators: Insights from nonequilibrium dynamical
  mean-field theory},\ }\href {https://doi.org/10.1103/PhysRevB.106.165106}
  {\bibfield  {journal} {\bibinfo  {journal} {Phys. Rev. B}\ }\textbf {\bibinfo
  {volume} {106}},\ \bibinfo {pages} {165106} (\bibinfo {year}
  {2022})}\BibitemShut {NoStop}%
\bibitem [{\citenamefont {Hedin}(1965)}]{hedin1965}%
  \BibitemOpen
  \bibfield  {author} {\bibinfo {author} {\bibfnamefont {L.}~\bibnamefont
  {Hedin}},\ }\bibfield  {title} {\bibinfo {title} {New method for calculating
  the one-particle green's function with application to the electron-gas
  problem},\ }\href {https://doi.org/10.1103/PhysRev.139.A796} {\bibfield
  {journal} {\bibinfo  {journal} {Phys. Rev.}\ }\textbf {\bibinfo {volume}
  {139}},\ \bibinfo {pages} {A796} (\bibinfo {year} {1965})}\BibitemShut
  {NoStop}%
\bibitem [{\citenamefont {Sun}\ and\ \citenamefont {Kotliar}(2002)}]{sun2002}%
  \BibitemOpen
  \bibfield  {author} {\bibinfo {author} {\bibfnamefont {P.}~\bibnamefont
  {Sun}}\ and\ \bibinfo {author} {\bibfnamefont {G.}~\bibnamefont {Kotliar}},\
  }\bibfield  {title} {\bibinfo {title} {Extended dynamical mean-field theory
  and $\mathrm{GW}$ method},\ }\href
  {https://doi.org/10.1103/PhysRevB.66.085120} {\bibfield  {journal} {\bibinfo
  {journal} {Phys. Rev. B}\ }\textbf {\bibinfo {volume} {66}},\ \bibinfo
  {pages} {085120} (\bibinfo {year} {2002})}\BibitemShut {NoStop}%
\bibitem [{\citenamefont {Ayral}\ \emph {et~al.}(2013)\citenamefont {Ayral},
  \citenamefont {Biermann},\ and\ \citenamefont {Werner}}]{ayral2013}%
  \BibitemOpen
  \bibfield  {author} {\bibinfo {author} {\bibfnamefont {T.}~\bibnamefont
  {Ayral}}, \bibinfo {author} {\bibfnamefont {S.}~\bibnamefont {Biermann}},\
  and\ \bibinfo {author} {\bibfnamefont {P.}~\bibnamefont {Werner}},\
  }\bibfield  {title} {\bibinfo {title} {Screening and nonlocal correlations in
  the extended hubbard model from self-consistent combined gw and dynamical
  mean field theory},\ }\href {https://doi.org/10.1103/PhysRevB.87.125149}
  {\bibfield  {journal} {\bibinfo  {journal} {Phys. Rev. B}\ }\textbf {\bibinfo
  {volume} {87}},\ \bibinfo {pages} {125149} (\bibinfo {year}
  {2013})}\BibitemShut {NoStop}%
\bibitem [{\citenamefont {Gole\ifmmode~\check{z}\else \v{z}\fi{}}\ \emph
  {et~al.}(2015)\citenamefont {Gole\ifmmode~\check{z}\else \v{z}\fi{}},
  \citenamefont {Eckstein},\ and\ \citenamefont {Werner}}]{golez2015}%
  \BibitemOpen
  \bibfield  {author} {\bibinfo {author} {\bibfnamefont {D.}~\bibnamefont
  {Gole\ifmmode~\check{z}\else \v{z}\fi{}}}, \bibinfo {author} {\bibfnamefont
  {M.}~\bibnamefont {Eckstein}},\ and\ \bibinfo {author} {\bibfnamefont
  {P.}~\bibnamefont {Werner}},\ }\bibfield  {title} {\bibinfo {title} {Dynamics
  of screening in photodoped mott insulators},\ }\href
  {https://doi.org/10.1103/PhysRevB.92.195123} {\bibfield  {journal} {\bibinfo
  {journal} {Phys. Rev. B}\ }\textbf {\bibinfo {volume} {92}},\ \bibinfo
  {pages} {195123} (\bibinfo {year} {2015})}\BibitemShut {NoStop}%
\bibitem [{\citenamefont {Nilsson}\ \emph {et~al.}(2017)\citenamefont
  {Nilsson}, \citenamefont {Boehnke}, \citenamefont {Werner},\ and\
  \citenamefont {Aryasetiawan}}]{nilsson2017}%
  \BibitemOpen
  \bibfield  {author} {\bibinfo {author} {\bibfnamefont {F.}~\bibnamefont
  {Nilsson}}, \bibinfo {author} {\bibfnamefont {L.}~\bibnamefont {Boehnke}},
  \bibinfo {author} {\bibfnamefont {P.}~\bibnamefont {Werner}},\ and\ \bibinfo
  {author} {\bibfnamefont {F.}~\bibnamefont {Aryasetiawan}},\ }\bibfield
  {title} {\bibinfo {title} {Multitier self-consistent $gw+\text{EDMFT}$},\
  }\href {https://doi.org/10.1103/PhysRevMaterials.1.043803} {\bibfield
  {journal} {\bibinfo  {journal} {Phys. Rev. Materials}\ }\textbf {\bibinfo
  {volume} {1}},\ \bibinfo {pages} {043803} (\bibinfo {year}
  {2017})}\BibitemShut {NoStop}%
\bibitem [{\citenamefont {Boehnke}\ \emph {et~al.}(2016)\citenamefont
  {Boehnke}, \citenamefont {Nilsson}, \citenamefont {Aryasetiawan},\ and\
  \citenamefont {Werner}}]{boehnke2016}%
  \BibitemOpen
  \bibfield  {author} {\bibinfo {author} {\bibfnamefont {L.}~\bibnamefont
  {Boehnke}}, \bibinfo {author} {\bibfnamefont {F.}~\bibnamefont {Nilsson}},
  \bibinfo {author} {\bibfnamefont {F.}~\bibnamefont {Aryasetiawan}},\ and\
  \bibinfo {author} {\bibfnamefont {P.}~\bibnamefont {Werner}},\ }\bibfield
  {title} {\bibinfo {title} {When strong correlations become weak: Consistent
  merging of $gw$ and dmft},\ }\href
  {https://doi.org/10.1103/PhysRevB.94.201106} {\bibfield  {journal} {\bibinfo
  {journal} {Phys. Rev. B}\ }\textbf {\bibinfo {volume} {94}},\ \bibinfo
  {pages} {201106} (\bibinfo {year} {2016})}\BibitemShut {NoStop}%
\bibitem [{\citenamefont {Georges}\ \emph {et~al.}(1996)\citenamefont
  {Georges}, \citenamefont {Kotliar}, \citenamefont {Krauth},\ and\
  \citenamefont {Rozenberg}}]{georges1996}%
  \BibitemOpen
  \bibfield  {author} {\bibinfo {author} {\bibfnamefont {A.}~\bibnamefont
  {Georges}}, \bibinfo {author} {\bibfnamefont {G.}~\bibnamefont {Kotliar}},
  \bibinfo {author} {\bibfnamefont {W.}~\bibnamefont {Krauth}},\ and\ \bibinfo
  {author} {\bibfnamefont {M.~J.}\ \bibnamefont {Rozenberg}},\ }\bibfield
  {title} {\bibinfo {title} {Dynamical mean-field theory of strongly correlated
  fermion systems and the limit of infinite dimensions},\ }\href
  {https://doi.org/10.1103/RevModPhys.68.13} {\bibfield  {journal} {\bibinfo
  {journal} {Rev. Mod. Phys.}\ }\textbf {\bibinfo {volume} {68}},\ \bibinfo
  {pages} {13} (\bibinfo {year} {1996})}\BibitemShut {NoStop}%
\bibitem [{\citenamefont {Cornaglia}\ and\ \citenamefont
  {Georges}(2007)}]{cornaglia2007}%
  \BibitemOpen
  \bibfield  {author} {\bibinfo {author} {\bibfnamefont {P.}~\bibnamefont
  {Cornaglia}}\ and\ \bibinfo {author} {\bibfnamefont {A.}~\bibnamefont
  {Georges}},\ }\bibfield  {title} {\bibinfo {title} {Theory of core-level
  photoemission and the x-ray edge singularity across the mott transition},\
  }\href@noop {} {\bibfield  {journal} {\bibinfo  {journal} {Physical Review
  B}\ }\textbf {\bibinfo {volume} {75}},\ \bibinfo {pages} {115112} (\bibinfo
  {year} {2007})}\BibitemShut {NoStop}%
\bibitem [{\citenamefont {Haverkort}\ \emph {et~al.}(2014)\citenamefont
  {Haverkort}, \citenamefont {Sangiovanni}, \citenamefont {Hansmann},
  \citenamefont {Toschi}, \citenamefont {Lu},\ and\ \citenamefont
  {Macke}}]{haverkort2014}%
  \BibitemOpen
  \bibfield  {author} {\bibinfo {author} {\bibfnamefont {M.}~\bibnamefont
  {Haverkort}}, \bibinfo {author} {\bibfnamefont {G.}~\bibnamefont
  {Sangiovanni}}, \bibinfo {author} {\bibfnamefont {P.}~\bibnamefont
  {Hansmann}}, \bibinfo {author} {\bibfnamefont {A.}~\bibnamefont {Toschi}},
  \bibinfo {author} {\bibfnamefont {Y.}~\bibnamefont {Lu}},\ and\ \bibinfo
  {author} {\bibfnamefont {S.}~\bibnamefont {Macke}},\ }\bibfield  {title}
  {\bibinfo {title} {Bands, resonances, edge singularities and excitons in core
  level spectroscopy investigated within the dynamical mean-field theory},\
  }\href@noop {} {\bibfield  {journal} {\bibinfo  {journal} {EPL (Europhysics
  Letters)}\ }\textbf {\bibinfo {volume} {108}},\ \bibinfo {pages} {57004}
  (\bibinfo {year} {2014})}\BibitemShut {NoStop}%
\bibitem [{\citenamefont {L{\"u}der}\ \emph {et~al.}(2017)\citenamefont
  {L{\"u}der}, \citenamefont {Sch{\"o}tt}, \citenamefont {Brena}, \citenamefont
  {Haverkort}, \citenamefont {Thunstr{\"o}m}, \citenamefont {Eriksson},
  \citenamefont {Sanyal}, \citenamefont {Di~Marco},\ and\ \citenamefont
  {Kvashnin}}]{luder2017}%
  \BibitemOpen
  \bibfield  {author} {\bibinfo {author} {\bibfnamefont {J.}~\bibnamefont
  {L{\"u}der}}, \bibinfo {author} {\bibfnamefont {J.}~\bibnamefont
  {Sch{\"o}tt}}, \bibinfo {author} {\bibfnamefont {B.}~\bibnamefont {Brena}},
  \bibinfo {author} {\bibfnamefont {M.~W.}\ \bibnamefont {Haverkort}}, \bibinfo
  {author} {\bibfnamefont {P.}~\bibnamefont {Thunstr{\"o}m}}, \bibinfo {author}
  {\bibfnamefont {O.}~\bibnamefont {Eriksson}}, \bibinfo {author}
  {\bibfnamefont {B.}~\bibnamefont {Sanyal}}, \bibinfo {author} {\bibfnamefont
  {I.}~\bibnamefont {Di~Marco}},\ and\ \bibinfo {author} {\bibfnamefont
  {Y.~O.}\ \bibnamefont {Kvashnin}},\ }\bibfield  {title} {\bibinfo {title}
  {Theory of l-edge spectroscopy of strongly correlated systems},\ }\href@noop
  {} {\bibfield  {journal} {\bibinfo  {journal} {Physical Review B}\ }\textbf
  {\bibinfo {volume} {96}},\ \bibinfo {pages} {245131} (\bibinfo {year}
  {2017})}\BibitemShut {NoStop}%
\bibitem [{\citenamefont {Hariki}\ \emph {et~al.}(2018)\citenamefont {Hariki},
  \citenamefont {Winder},\ and\ \citenamefont {Kune{\v{s}}}}]{hariki2018}%
  \BibitemOpen
  \bibfield  {author} {\bibinfo {author} {\bibfnamefont {A.}~\bibnamefont
  {Hariki}}, \bibinfo {author} {\bibfnamefont {M.}~\bibnamefont {Winder}},\
  and\ \bibinfo {author} {\bibfnamefont {J.}~\bibnamefont {Kune{\v{s}}}},\
  }\bibfield  {title} {\bibinfo {title} {Continuum charge excitations in
  high-valence transition-metal oxides revealed by resonant inelastic x-ray
  scattering},\ }\href@noop {} {\bibfield  {journal} {\bibinfo  {journal}
  {Physical review letters}\ }\textbf {\bibinfo {volume} {121}},\ \bibinfo
  {pages} {126403} (\bibinfo {year} {2018})}\BibitemShut {NoStop}%
\bibitem [{\citenamefont {Tanaka}\ and\ \citenamefont {Jo}(1994)}]{tanaka1994}%
  \BibitemOpen
  \bibfield  {author} {\bibinfo {author} {\bibfnamefont {A.}~\bibnamefont
  {Tanaka}}\ and\ \bibinfo {author} {\bibfnamefont {T.}~\bibnamefont {Jo}},\
  }\bibfield  {title} {\bibinfo {title} {Resonant 3d, 3pand 3sphotoemission in
  transition metal oxides predicted at 2pthreshold},\ }\href
  {https://doi.org/10.1143/jpsj.63.2788} {\bibfield  {journal} {\bibinfo
  {journal} {Journal of the Physical Society of Japan}\ }\textbf {\bibinfo
  {volume} {63}},\ \bibinfo {pages} {2788–2807} (\bibinfo {year}
  {1994})}\BibitemShut {NoStop}%
\bibitem [{\citenamefont {Haverkort}\ \emph {et~al.}(2012)\citenamefont
  {Haverkort}, \citenamefont {Zwierzycki},\ and\ \citenamefont
  {Andersen}}]{haverkort2012}%
  \BibitemOpen
  \bibfield  {author} {\bibinfo {author} {\bibfnamefont {M.~W.}\ \bibnamefont
  {Haverkort}}, \bibinfo {author} {\bibfnamefont {M.}~\bibnamefont
  {Zwierzycki}},\ and\ \bibinfo {author} {\bibfnamefont {O.~K.}\ \bibnamefont
  {Andersen}},\ }\bibfield  {title} {\bibinfo {title} {Multiplet ligand-field
  theory using wannier orbitals},\ }\href
  {https://doi.org/10.1103/PhysRevB.85.165113} {\bibfield  {journal} {\bibinfo
  {journal} {Phys. Rev. B}\ }\textbf {\bibinfo {volume} {85}},\ \bibinfo
  {pages} {165113} (\bibinfo {year} {2012})}\BibitemShut {NoStop}%
\bibitem [{\citenamefont {\ifmmode~\check{S}\else \v{S}\fi{}ipr}\ \emph
  {et~al.}(2011)\citenamefont {\ifmmode~\check{S}\else \v{S}\fi{}ipr},
  \citenamefont {Min\'ar}, \citenamefont {Scherz}, \citenamefont {Wende},\ and\
  \citenamefont {Ebert}}]{sipr2011}%
  \BibitemOpen
  \bibfield  {author} {\bibinfo {author} {\bibfnamefont {O.}~\bibnamefont
  {\ifmmode~\check{S}\else \v{S}\fi{}ipr}}, \bibinfo {author} {\bibfnamefont
  {J.}~\bibnamefont {Min\'ar}}, \bibinfo {author} {\bibfnamefont
  {A.}~\bibnamefont {Scherz}}, \bibinfo {author} {\bibfnamefont
  {H.}~\bibnamefont {Wende}},\ and\ \bibinfo {author} {\bibfnamefont
  {H.}~\bibnamefont {Ebert}},\ }\bibfield  {title} {\bibinfo {title} {Many-body
  effects in x-ray absorption and magnetic circular dichroism spectra within
  the lsda+dmft framework},\ }\href
  {https://doi.org/10.1103/PhysRevB.84.115102} {\bibfield  {journal} {\bibinfo
  {journal} {Phys. Rev. B}\ }\textbf {\bibinfo {volume} {84}},\ \bibinfo
  {pages} {115102} (\bibinfo {year} {2011})}\BibitemShut {NoStop}%
\bibitem [{\citenamefont {Aoki}\ \emph {et~al.}(2014)\citenamefont {Aoki},
  \citenamefont {Tsuji}, \citenamefont {Eckstein}, \citenamefont {Kollar},
  \citenamefont {Oka},\ and\ \citenamefont {Werner}}]{aoki2014_rev}%
  \BibitemOpen
  \bibfield  {author} {\bibinfo {author} {\bibfnamefont {H.}~\bibnamefont
  {Aoki}}, \bibinfo {author} {\bibfnamefont {N.}~\bibnamefont {Tsuji}},
  \bibinfo {author} {\bibfnamefont {M.}~\bibnamefont {Eckstein}}, \bibinfo
  {author} {\bibfnamefont {M.}~\bibnamefont {Kollar}}, \bibinfo {author}
  {\bibfnamefont {T.}~\bibnamefont {Oka}},\ and\ \bibinfo {author}
  {\bibfnamefont {P.}~\bibnamefont {Werner}},\ }\bibfield  {title} {\bibinfo
  {title} {Nonequilibrium dynamical mean-field theory and its applications},\
  }\href {http://link.aps.org/doi/10.1103/RevModPhys.86.779} {\bibfield
  {journal} {\bibinfo  {journal} {Rev. Mod. Phys.}\ }\textbf {\bibinfo {volume}
  {86}},\ \bibinfo {pages} {779} (\bibinfo {year} {2014})}\BibitemShut
  {NoStop}%
\bibitem [{\citenamefont {Erpenbeck}\ \emph {et~al.}(2023)\citenamefont
  {Erpenbeck}, \citenamefont {Gull},\ and\ \citenamefont
  {Cohen}}]{erpenbeck2023}%
  \BibitemOpen
  \bibfield  {author} {\bibinfo {author} {\bibfnamefont {A.}~\bibnamefont
  {Erpenbeck}}, \bibinfo {author} {\bibfnamefont {E.}~\bibnamefont {Gull}},\
  and\ \bibinfo {author} {\bibfnamefont {G.}~\bibnamefont {Cohen}},\ }\bibfield
   {title} {\bibinfo {title} {Quantum monte carlo method in the steady state},\
  }\href {https://doi.org/10.1103/PhysRevLett.130.186301} {\bibfield  {journal}
  {\bibinfo  {journal} {Phys. Rev. Lett.}\ }\textbf {\bibinfo {volume} {130}},\
  \bibinfo {pages} {186301} (\bibinfo {year} {2023})}\BibitemShut {NoStop}%
\bibitem [{\citenamefont {K\"unzel}\ \emph {et~al.}(2024)\citenamefont
  {K\"unzel}, \citenamefont {Erpenbeck}, \citenamefont {Werner}, \citenamefont
  {Arrigoni}, \citenamefont {Gull}, \citenamefont {Cohen},\ and\ \citenamefont
  {Eckstein}}]{kunzel2024}%
  \BibitemOpen
  \bibfield  {author} {\bibinfo {author} {\bibfnamefont {F.}~\bibnamefont
  {K\"unzel}}, \bibinfo {author} {\bibfnamefont {A.}~\bibnamefont {Erpenbeck}},
  \bibinfo {author} {\bibfnamefont {D.}~\bibnamefont {Werner}}, \bibinfo
  {author} {\bibfnamefont {E.}~\bibnamefont {Arrigoni}}, \bibinfo {author}
  {\bibfnamefont {E.}~\bibnamefont {Gull}}, \bibinfo {author} {\bibfnamefont
  {G.}~\bibnamefont {Cohen}},\ and\ \bibinfo {author} {\bibfnamefont
  {M.}~\bibnamefont {Eckstein}},\ }\bibfield  {title} {\bibinfo {title}
  {Numerically exact simulation of photodoped mott insulators},\ }\href
  {https://doi.org/10.1103/PhysRevLett.132.176501} {\bibfield  {journal}
  {\bibinfo  {journal} {Phys. Rev. Lett.}\ }\textbf {\bibinfo {volume} {132}},\
  \bibinfo {pages} {176501} (\bibinfo {year} {2024})}\BibitemShut {NoStop}%
\bibitem [{\citenamefont {Emery}(1987)}]{emery1987}%
  \BibitemOpen
  \bibfield  {author} {\bibinfo {author} {\bibfnamefont {V.}~\bibnamefont
  {Emery}},\ }\bibfield  {title} {\bibinfo {title} {Theory of high-t c
  superconductivity in oxides},\ }\href@noop {} {\bibfield  {journal} {\bibinfo
   {journal} {Phys. Rev. Lett.}\ }\textbf {\bibinfo {volume} {58}},\ \bibinfo
  {pages} {2794} (\bibinfo {year} {1987})}\BibitemShut {NoStop}%
\bibitem [{\citenamefont {Gole\ifmmode~\check{z}\else \v{z}\fi{}}\ \emph
  {et~al.}(2022)\citenamefont {Gole\ifmmode~\check{z}\else \v{z}\fi{}},
  \citenamefont {Dufresne}, \citenamefont {Kim}, \citenamefont {Boschini},
  \citenamefont {Chu}, \citenamefont {Murakami}, \citenamefont {Levy},
  \citenamefont {Mills}, \citenamefont {Zhdanovich}, \citenamefont {Isobe},
  \citenamefont {Takagi}, \citenamefont {Kaiser}, \citenamefont {Werner},
  \citenamefont {Jones}, \citenamefont {Georges}, \citenamefont {Damascelli},\
  and\ \citenamefont {Millis}}]{golez2022}%
  \BibitemOpen
  \bibfield  {author} {\bibinfo {author} {\bibfnamefont {D.}~\bibnamefont
  {Gole\ifmmode~\check{z}\else \v{z}\fi{}}}, \bibinfo {author} {\bibfnamefont
  {S.~K.~Y.}\ \bibnamefont {Dufresne}}, \bibinfo {author} {\bibfnamefont
  {M.-J.}\ \bibnamefont {Kim}}, \bibinfo {author} {\bibfnamefont
  {F.}~\bibnamefont {Boschini}}, \bibinfo {author} {\bibfnamefont
  {H.}~\bibnamefont {Chu}}, \bibinfo {author} {\bibfnamefont {Y.}~\bibnamefont
  {Murakami}}, \bibinfo {author} {\bibfnamefont {G.}~\bibnamefont {Levy}},
  \bibinfo {author} {\bibfnamefont {A.~K.}\ \bibnamefont {Mills}}, \bibinfo
  {author} {\bibfnamefont {S.}~\bibnamefont {Zhdanovich}}, \bibinfo {author}
  {\bibfnamefont {M.}~\bibnamefont {Isobe}}, \bibinfo {author} {\bibfnamefont
  {H.}~\bibnamefont {Takagi}}, \bibinfo {author} {\bibfnamefont
  {S.}~\bibnamefont {Kaiser}}, \bibinfo {author} {\bibfnamefont
  {P.}~\bibnamefont {Werner}}, \bibinfo {author} {\bibfnamefont {D.~J.}\
  \bibnamefont {Jones}}, \bibinfo {author} {\bibfnamefont {A.}~\bibnamefont
  {Georges}}, \bibinfo {author} {\bibfnamefont {A.}~\bibnamefont
  {Damascelli}},\ and\ \bibinfo {author} {\bibfnamefont {A.~J.}\ \bibnamefont
  {Millis}},\ }\bibfield  {title} {\bibinfo {title} {Unveiling the underlying
  interactions in ${\mathrm{ta}}_{2}{\mathrm{nise}}_{5}$ from photoinduced
  lifetime change},\ }\href {https://doi.org/10.1103/PhysRevB.106.L121106}
  {\bibfield  {journal} {\bibinfo  {journal} {Phys. Rev. B}\ }\textbf {\bibinfo
  {volume} {106}},\ \bibinfo {pages} {L121106} (\bibinfo {year}
  {2022})}\BibitemShut {NoStop}%
\bibitem [{\citenamefont {Gole\ifmmode~\check{z}\else \v{z}\fi{}}\ \emph
  {et~al.}(2017)\citenamefont {Gole\ifmmode~\check{z}\else \v{z}\fi{}},
  \citenamefont {Boehnke}, \citenamefont {Strand}, \citenamefont {Eckstein},\
  and\ \citenamefont {Werner}}]{golez2017}%
  \BibitemOpen
  \bibfield  {author} {\bibinfo {author} {\bibfnamefont {D.}~\bibnamefont
  {Gole\ifmmode~\check{z}\else \v{z}\fi{}}}, \bibinfo {author} {\bibfnamefont
  {L.}~\bibnamefont {Boehnke}}, \bibinfo {author} {\bibfnamefont {H.~U.~R.}\
  \bibnamefont {Strand}}, \bibinfo {author} {\bibfnamefont {M.}~\bibnamefont
  {Eckstein}},\ and\ \bibinfo {author} {\bibfnamefont {P.}~\bibnamefont
  {Werner}},\ }\bibfield  {title} {\bibinfo {title} {Nonequilibrium
  $gw+\mathrm{EDMFT}$: Antiscreening and inverted populations from nonlocal
  correlations},\ }\href {https://doi.org/10.1103/PhysRevLett.118.246402}
  {\bibfield  {journal} {\bibinfo  {journal} {Phys. Rev. Lett.}\ }\textbf
  {\bibinfo {volume} {118}},\ \bibinfo {pages} {246402} (\bibinfo {year}
  {2017})}\BibitemShut {NoStop}%
\bibitem [{\citenamefont {Lee}\ and\ \citenamefont {Haule}(2017)}]{lee2017}%
  \BibitemOpen
  \bibfield  {author} {\bibinfo {author} {\bibfnamefont {J.}~\bibnamefont
  {Lee}}\ and\ \bibinfo {author} {\bibfnamefont {K.}~\bibnamefont {Haule}},\
  }\bibfield  {title} {\bibinfo {title} {Diatomic molecule as a testbed for
  combining dmft with electronic structure methods such as $gw$ and dft},\
  }\href {https://doi.org/10.1103/PhysRevB.95.155104} {\bibfield  {journal}
  {\bibinfo  {journal} {Phys. Rev. B}\ }\textbf {\bibinfo {volume} {95}},\
  \bibinfo {pages} {155104} (\bibinfo {year} {2017})}\BibitemShut {NoStop}%
\bibitem [{\citenamefont {Vu\ifmmode \check{c}\else \v{c}\fi{}i\ifmmode
  \check{c}\else \v{c}\fi{}evi\ifmmode~\acute{c}\else \'{c}\fi{}}\ \emph
  {et~al.}(2018)\citenamefont {Vu\ifmmode \check{c}\else \v{c}\fi{}i\ifmmode
  \check{c}\else \v{c}\fi{}evi\ifmmode~\acute{c}\else \'{c}\fi{}},
  \citenamefont {Wentzell}, \citenamefont {Ferrero},\ and\ \citenamefont
  {Parcollet}}]{vucicevic2018}%
  \BibitemOpen
  \bibfield  {author} {\bibinfo {author} {\bibfnamefont {J.}~\bibnamefont
  {Vu\ifmmode \check{c}\else \v{c}\fi{}i\ifmmode \check{c}\else
  \v{c}\fi{}evi\ifmmode~\acute{c}\else \'{c}\fi{}}}, \bibinfo {author}
  {\bibfnamefont {N.}~\bibnamefont {Wentzell}}, \bibinfo {author}
  {\bibfnamefont {M.}~\bibnamefont {Ferrero}},\ and\ \bibinfo {author}
  {\bibfnamefont {O.}~\bibnamefont {Parcollet}},\ }\bibfield  {title} {\bibinfo
  {title} {Practical consequences of the luttinger-ward functional
  multivaluedness for cluster dmft methods},\ }\href
  {https://doi.org/10.1103/PhysRevB.97.125141} {\bibfield  {journal} {\bibinfo
  {journal} {Phys. Rev. B}\ }\textbf {\bibinfo {volume} {97}},\ \bibinfo
  {pages} {125141} (\bibinfo {year} {2018})}\BibitemShut {NoStop}%
\bibitem [{\citenamefont {Backes}\ \emph {et~al.}(2022)\citenamefont {Backes},
  \citenamefont {Sim},\ and\ \citenamefont {Biermann}}]{backes2022}%
  \BibitemOpen
  \bibfield  {author} {\bibinfo {author} {\bibfnamefont {S.}~\bibnamefont
  {Backes}}, \bibinfo {author} {\bibfnamefont {J.-H.}\ \bibnamefont {Sim}},\
  and\ \bibinfo {author} {\bibfnamefont {S.}~\bibnamefont {Biermann}},\
  }\bibfield  {title} {\bibinfo {title} {Nonlocal correlation effects in
  fermionic many-body systems: Overcoming the noncausality problem},\ }\href
  {https://doi.org/10.1103/PhysRevB.105.245115} {\bibfield  {journal} {\bibinfo
   {journal} {Phys. Rev. B}\ }\textbf {\bibinfo {volume} {105}},\ \bibinfo
  {pages} {245115} (\bibinfo {year} {2022})}\BibitemShut {NoStop}%
\bibitem [{\citenamefont {Chen}\ \emph {et~al.}(2022)\citenamefont {Chen},
  \citenamefont {Petocchi},\ and\ \citenamefont {Werner}}]{chen2022}%
  \BibitemOpen
  \bibfield  {author} {\bibinfo {author} {\bibfnamefont {J.}~\bibnamefont
  {Chen}}, \bibinfo {author} {\bibfnamefont {F.}~\bibnamefont {Petocchi}},\
  and\ \bibinfo {author} {\bibfnamefont {P.}~\bibnamefont {Werner}},\
  }\bibfield  {title} {\bibinfo {title} {Causal versus local
  $gw+\mathrm{EDMFT}$ scheme and application to the triangular-lattice extended
  hubbard model},\ }\href {https://doi.org/10.1103/PhysRevB.105.085102}
  {\bibfield  {journal} {\bibinfo  {journal} {Phys. Rev. B}\ }\textbf {\bibinfo
  {volume} {105}},\ \bibinfo {pages} {085102} (\bibinfo {year}
  {2022})}\BibitemShut {NoStop}%
\bibitem [{\citenamefont {van~der Laan}\ \emph {et~al.}(1986)\citenamefont
  {van~der Laan}, \citenamefont {Zaanen}, \citenamefont {Sawatzky},
  \citenamefont {Karnatak},\ and\ \citenamefont {Esteva}}]{laan1986}%
  \BibitemOpen
  \bibfield  {author} {\bibinfo {author} {\bibfnamefont {G.}~\bibnamefont
  {van~der Laan}}, \bibinfo {author} {\bibfnamefont {J.}~\bibnamefont
  {Zaanen}}, \bibinfo {author} {\bibfnamefont {G.~A.}\ \bibnamefont
  {Sawatzky}}, \bibinfo {author} {\bibfnamefont {R.}~\bibnamefont {Karnatak}},\
  and\ \bibinfo {author} {\bibfnamefont {J.-M.}\ \bibnamefont {Esteva}},\
  }\bibfield  {title} {\bibinfo {title} {Comparison of x-ray absorption with
  x-ray photoemission of nickel dihalides and nio},\ }\href
  {https://doi.org/10.1103/PhysRevB.33.4253} {\bibfield  {journal} {\bibinfo
  {journal} {Phys. Rev. B}\ }\textbf {\bibinfo {volume} {33}},\ \bibinfo
  {pages} {4253} (\bibinfo {year} {1986})}\BibitemShut {NoStop}%
\bibitem [{\citenamefont {Hariki}\ \emph {et~al.}(2017)\citenamefont {Hariki},
  \citenamefont {Uozumi},\ and\ \citenamefont {Kune\ifmmode~\check{s}\else
  \v{s}\fi{}}}]{hariki2017}%
  \BibitemOpen
  \bibfield  {author} {\bibinfo {author} {\bibfnamefont {A.}~\bibnamefont
  {Hariki}}, \bibinfo {author} {\bibfnamefont {T.}~\bibnamefont {Uozumi}},\
  and\ \bibinfo {author} {\bibfnamefont {J.}~\bibnamefont
  {Kune\ifmmode~\check{s}\else \v{s}\fi{}}},\ }\bibfield  {title} {\bibinfo
  {title} {Lda+dmft approach to core-level spectroscopy: Application to $3d$
  transition metal compounds},\ }\href
  {https://doi.org/10.1103/PhysRevB.96.045111} {\bibfield  {journal} {\bibinfo
  {journal} {Phys. Rev. B}\ }\textbf {\bibinfo {volume} {96}},\ \bibinfo
  {pages} {045111} (\bibinfo {year} {2017})}\BibitemShut {NoStop}%
\bibitem [{\citenamefont {Zaanen}\ \emph {et~al.}(1986)\citenamefont {Zaanen},
  \citenamefont {Westra},\ and\ \citenamefont {Sawatzky}}]{zaanen1986}%
  \BibitemOpen
  \bibfield  {author} {\bibinfo {author} {\bibfnamefont {J.}~\bibnamefont
  {Zaanen}}, \bibinfo {author} {\bibfnamefont {C.}~\bibnamefont {Westra}},\
  and\ \bibinfo {author} {\bibfnamefont {G.~A.}\ \bibnamefont {Sawatzky}},\
  }\bibfield  {title} {\bibinfo {title} {Determination of the electronic
  structure of transition-metal compounds: 2p x-ray photoemission spectroscopy
  of the nickel dihalides},\ }\href {https://doi.org/10.1103/PhysRevB.33.8060}
  {\bibfield  {journal} {\bibinfo  {journal} {Phys. Rev. B}\ }\textbf {\bibinfo
  {volume} {33}},\ \bibinfo {pages} {8060} (\bibinfo {year}
  {1986})}\BibitemShut {NoStop}%
\bibitem [{\citenamefont {Zhang}\ and\ \citenamefont {Rice}(1988)}]{zhang1988}%
  \BibitemOpen
  \bibfield  {author} {\bibinfo {author} {\bibfnamefont {F.~C.}\ \bibnamefont
  {Zhang}}\ and\ \bibinfo {author} {\bibfnamefont {T.~M.}\ \bibnamefont
  {Rice}},\ }\bibfield  {title} {\bibinfo {title} {Effective hamiltonian for
  the superconducting {Cu} oxides},\ }\href@noop {} {\bibfield  {journal}
  {\bibinfo  {journal} {Phys. Rev. B}\ }\textbf {\bibinfo {volume} {37}},\
  \bibinfo {pages} {3759} (\bibinfo {year} {1988})}\BibitemShut {NoStop}%
\bibitem [{\citenamefont {Zaanen}\ and\ \citenamefont
  {Ole\ifmmode~\acute{s}\else \'{s}\fi{}}(1988)}]{zannen1988}%
  \BibitemOpen
  \bibfield  {author} {\bibinfo {author} {\bibfnamefont {J.}~\bibnamefont
  {Zaanen}}\ and\ \bibinfo {author} {\bibfnamefont {A.~M.}\ \bibnamefont
  {Ole\ifmmode~\acute{s}\else \'{s}\fi{}}},\ }\bibfield  {title} {\bibinfo
  {title} {Canonical perturbation theory and the two-band model for
  high-${T}_{c}$ superconductors},\ }\href
  {https://doi.org/10.1103/PhysRevB.37.9423} {\bibfield  {journal} {\bibinfo
  {journal} {Phys. Rev. B}\ }\textbf {\bibinfo {volume} {37}},\ \bibinfo
  {pages} {9423} (\bibinfo {year} {1988})}\BibitemShut {NoStop}%
\bibitem [{\citenamefont {Ramak}\ and\ \citenamefont
  {Prelovek}(1989)}]{ramsak1989}%
  \BibitemOpen
  \bibfield  {author} {\bibinfo {author} {\bibfnamefont {A.}~\bibnamefont
  {Ramak}}\ and\ \bibinfo {author} {\bibfnamefont {P.}~\bibnamefont
  {Prelovek}},\ }\bibfield  {title} {\bibinfo {title} {Comparison of effective
  models for ${\mathrm{cuo}}_{2}$ layers in oxide superconductors},\ }\href
  {https://doi.org/10.1103/PhysRevB.40.2239} {\bibfield  {journal} {\bibinfo
  {journal} {Phys. Rev. B}\ }\textbf {\bibinfo {volume} {40}},\ \bibinfo
  {pages} {2239} (\bibinfo {year} {1989})}\BibitemShut {NoStop}%
\bibitem [{\citenamefont {Feiner}\ \emph {et~al.}(1996)\citenamefont {Feiner},
  \citenamefont {Jefferson},\ and\ \citenamefont {Raimondi}}]{feiner1996}%
  \BibitemOpen
  \bibfield  {author} {\bibinfo {author} {\bibfnamefont {L.~F.}\ \bibnamefont
  {Feiner}}, \bibinfo {author} {\bibfnamefont {J.~H.}\ \bibnamefont
  {Jefferson}},\ and\ \bibinfo {author} {\bibfnamefont {R.}~\bibnamefont
  {Raimondi}},\ }\bibfield  {title} {\bibinfo {title} {Effective single-band
  models for the high-${\mathit{t}}_{\mathit{c}}$ cuprates. i. coulomb
  interactions},\ }\href {https://doi.org/10.1103/PhysRevB.53.8751} {\bibfield
  {journal} {\bibinfo  {journal} {Phys. Rev. B}\ }\textbf {\bibinfo {volume}
  {53}},\ \bibinfo {pages} {8751} (\bibinfo {year} {1996})}\BibitemShut
  {NoStop}%
\bibitem [{\citenamefont {Jefferson}\ \emph {et~al.}(1992)\citenamefont
  {Jefferson}, \citenamefont {Eskes},\ and\ \citenamefont
  {Feiner}}]{Jefferson1992}%
  \BibitemOpen
  \bibfield  {author} {\bibinfo {author} {\bibfnamefont {J.~H.}\ \bibnamefont
  {Jefferson}}, \bibinfo {author} {\bibfnamefont {H.}~\bibnamefont {Eskes}},\
  and\ \bibinfo {author} {\bibfnamefont {L.~F.}\ \bibnamefont {Feiner}},\
  }\bibfield  {title} {\bibinfo {title} {Derivation of a single-band model for
  ${\mathrm{cuo}}_{2}$ planes by a cell-perturbation method},\ }\href
  {https://doi.org/10.1103/PhysRevB.45.7959} {\bibfield  {journal} {\bibinfo
  {journal} {Phys. Rev. B}\ }\textbf {\bibinfo {volume} {45}},\ \bibinfo
  {pages} {7959} (\bibinfo {year} {1992})}\BibitemShut {NoStop}%
\bibitem [{\citenamefont {Smith}(1988)}]{smith1988}%
  \BibitemOpen
  \bibfield  {author} {\bibinfo {author} {\bibfnamefont {N.~V.}\ \bibnamefont
  {Smith}},\ }\bibfield  {title} {\bibinfo {title} {Inverse photoemission},\
  }\href {https://doi.org/10.1088/0034-4885/51/9/003} {\bibfield  {journal}
  {\bibinfo  {journal} {Reports on Progress in Physics}\ }\textbf {\bibinfo
  {volume} {51}},\ \bibinfo {pages} {1227} (\bibinfo {year}
  {1988})}\BibitemShut {NoStop}%
\bibitem [{\citenamefont {Gillmeister}\ \emph {et~al.}(2020)\citenamefont
  {Gillmeister}, \citenamefont {Gole{\v{z}}}, \citenamefont {Chiang},
  \citenamefont {Bittner}, \citenamefont {Pavlyukh}, \citenamefont {Berakdar},
  \citenamefont {Werner},\ and\ \citenamefont {Widdra}}]{gillmeister2020}%
  \BibitemOpen
  \bibfield  {author} {\bibinfo {author} {\bibfnamefont {K.}~\bibnamefont
  {Gillmeister}}, \bibinfo {author} {\bibfnamefont {D.}~\bibnamefont
  {Gole{\v{z}}}}, \bibinfo {author} {\bibfnamefont {C.-T.}\ \bibnamefont
  {Chiang}}, \bibinfo {author} {\bibfnamefont {N.}~\bibnamefont {Bittner}},
  \bibinfo {author} {\bibfnamefont {Y.}~\bibnamefont {Pavlyukh}}, \bibinfo
  {author} {\bibfnamefont {J.}~\bibnamefont {Berakdar}}, \bibinfo {author}
  {\bibfnamefont {P.}~\bibnamefont {Werner}},\ and\ \bibinfo {author}
  {\bibfnamefont {W.}~\bibnamefont {Widdra}},\ }\bibfield  {title} {\bibinfo
  {title} {Ultrafast coupled charge and spin dynamics in strongly correlated
  nio},\ }\href@noop {} {\bibfield  {journal} {\bibinfo  {journal} {Nature
  communications}\ }\textbf {\bibinfo {volume} {11}},\ \bibinfo {pages} {4095}
  (\bibinfo {year} {2020})}\BibitemShut {NoStop}%
\bibitem [{\citenamefont {Petek}\ and\ \citenamefont
  {Ogawa}(1997)}]{petek1997}%
  \BibitemOpen
  \bibfield  {author} {\bibinfo {author} {\bibfnamefont {H.}~\bibnamefont
  {Petek}}\ and\ \bibinfo {author} {\bibfnamefont {S.}~\bibnamefont {Ogawa}},\
  }\bibfield  {title} {\bibinfo {title} {Femtosecond time-resolved two-photon
  photoemission studies of electron dynamics in metals},\ }\href
  {https://doi.org/https://doi.org/10.1016/S0079-6816(98)00002-1} {\bibfield
  {journal} {\bibinfo  {journal} {Progress in Surface Science}\ }\textbf
  {\bibinfo {volume} {56}},\ \bibinfo {pages} {239} (\bibinfo {year}
  {1997})}\BibitemShut {NoStop}%
\bibitem [{\citenamefont {Okada}\ and\ \citenamefont
  {Kotani}(1997)}]{okada1997intersite}%
  \BibitemOpen
  \bibfield  {author} {\bibinfo {author} {\bibfnamefont {K.}~\bibnamefont
  {Okada}}\ and\ \bibinfo {author} {\bibfnamefont {A.}~\bibnamefont {Kotani}},\
  }\bibfield  {title} {\bibinfo {title} {Intersite coulomb interactions in
  quasi-one-dimensional copper oxides},\ }\href@noop {} {\bibfield  {journal}
  {\bibinfo  {journal} {journal of the physical society of japan}\ }\textbf
  {\bibinfo {volume} {66}},\ \bibinfo {pages} {341} (\bibinfo {year}
  {1997})}\BibitemShut {NoStop}%
\bibitem [{\citenamefont {Matsuda}\ \emph {et~al.}(1994)\citenamefont
  {Matsuda}, \citenamefont {Hirabayashi}, \citenamefont {Kawamoto},
  \citenamefont {Nabatame}, \citenamefont {Tokizaki},\ and\ \citenamefont
  {Nakamura}}]{matsuda1994}%
  \BibitemOpen
  \bibfield  {author} {\bibinfo {author} {\bibfnamefont {K.}~\bibnamefont
  {Matsuda}}, \bibinfo {author} {\bibfnamefont {I.}~\bibnamefont
  {Hirabayashi}}, \bibinfo {author} {\bibfnamefont {K.}~\bibnamefont
  {Kawamoto}}, \bibinfo {author} {\bibfnamefont {T.}~\bibnamefont {Nabatame}},
  \bibinfo {author} {\bibfnamefont {T.}~\bibnamefont {Tokizaki}},\ and\
  \bibinfo {author} {\bibfnamefont {A.}~\bibnamefont {Nakamura}},\ }\bibfield
  {title} {\bibinfo {title} {Femtosecond spectroscopic studies of the ultrafast
  relaxation process in the charge-transfer state of insulating cuprates},\
  }\href {https://doi.org/10.1103/PhysRevB.50.4097} {\bibfield  {journal}
  {\bibinfo  {journal} {Phys. Rev. B}\ }\textbf {\bibinfo {volume} {50}},\
  \bibinfo {pages} {4097} (\bibinfo {year} {1994})}\BibitemShut {NoStop}%
\bibitem [{\citenamefont {Okamoto}\ \emph {et~al.}(2007)\citenamefont
  {Okamoto}, \citenamefont {Matsuzaki}, \citenamefont {Wakabayashi},
  \citenamefont {Takahashi},\ and\ \citenamefont {Hasegawa}}]{okamoto2007}%
  \BibitemOpen
  \bibfield  {author} {\bibinfo {author} {\bibfnamefont {H.}~\bibnamefont
  {Okamoto}}, \bibinfo {author} {\bibfnamefont {H.}~\bibnamefont {Matsuzaki}},
  \bibinfo {author} {\bibfnamefont {T.}~\bibnamefont {Wakabayashi}}, \bibinfo
  {author} {\bibfnamefont {Y.}~\bibnamefont {Takahashi}},\ and\ \bibinfo
  {author} {\bibfnamefont {T.}~\bibnamefont {Hasegawa}},\ }\bibfield  {title}
  {\bibinfo {title} {Photoinduced metallic state mediated by spin-charge
  separation in a one-dimensional organic mott insulator},\ }\href
  {https://doi.org/10.1103/PhysRevLett.98.037401} {\bibfield  {journal}
  {\bibinfo  {journal} {Phys. Rev. Lett.}\ }\textbf {\bibinfo {volume} {98}},\
  \bibinfo {pages} {037401} (\bibinfo {year} {2007})}\BibitemShut {NoStop}%
\bibitem [{\citenamefont {Okamoto}\ \emph {et~al.}(2011)\citenamefont
  {Okamoto}, \citenamefont {Miyagoe}, \citenamefont {Kobayashi}, \citenamefont
  {Uemura}, \citenamefont {Nishioka}, \citenamefont {Matsuzaki}, \citenamefont
  {Sawa},\ and\ \citenamefont {Tokura}}]{okamoto2011}%
  \BibitemOpen
  \bibfield  {author} {\bibinfo {author} {\bibfnamefont {H.}~\bibnamefont
  {Okamoto}}, \bibinfo {author} {\bibfnamefont {T.}~\bibnamefont {Miyagoe}},
  \bibinfo {author} {\bibfnamefont {K.}~\bibnamefont {Kobayashi}}, \bibinfo
  {author} {\bibfnamefont {H.}~\bibnamefont {Uemura}}, \bibinfo {author}
  {\bibfnamefont {H.}~\bibnamefont {Nishioka}}, \bibinfo {author}
  {\bibfnamefont {H.}~\bibnamefont {Matsuzaki}}, \bibinfo {author}
  {\bibfnamefont {A.}~\bibnamefont {Sawa}},\ and\ \bibinfo {author}
  {\bibfnamefont {Y.}~\bibnamefont {Tokura}},\ }\bibfield  {title} {\bibinfo
  {title} {Photoinduced transition from {M}ott insulator to metal in the
  undoped cuprates $\mathrm{Nd}_{2}\mathrm{CuO}_{4}$ and
  $\mathrm{La}_{2}\mathrm{CuO}_{4}$},\ }\href@noop {} {\bibfield  {journal}
  {\bibinfo  {journal} {Phys. Rev. B}\ }\textbf {\bibinfo {volume} {83}},\
  \bibinfo {pages} {125102} (\bibinfo {year} {2011})}\BibitemShut {NoStop}%
\bibitem [{\citenamefont {Novelli}\ \emph {et~al.}(2012)\citenamefont
  {Novelli}, \citenamefont {Fausti}, \citenamefont {Reul}, \citenamefont
  {Cilento}, \citenamefont {van Loosdrecht}, \citenamefont {Nugroho},
  \citenamefont {Palstra}, \citenamefont {Gr\"{u}ninger},\ and\ \citenamefont
  {Parmigiani}}]{novelli12}%
  \BibitemOpen
  \bibfield  {author} {\bibinfo {author} {\bibfnamefont {F.}~\bibnamefont
  {Novelli}}, \bibinfo {author} {\bibfnamefont {D.}~\bibnamefont {Fausti}},
  \bibinfo {author} {\bibfnamefont {J.}~\bibnamefont {Reul}}, \bibinfo {author}
  {\bibfnamefont {F.}~\bibnamefont {Cilento}}, \bibinfo {author} {\bibfnamefont
  {P.~H.~M.}\ \bibnamefont {van Loosdrecht}}, \bibinfo {author} {\bibfnamefont
  {A.~A.}\ \bibnamefont {Nugroho}}, \bibinfo {author} {\bibfnamefont
  {T.~T.~M.}\ \bibnamefont {Palstra}}, \bibinfo {author} {\bibfnamefont
  {M.}~\bibnamefont {Gr\"{u}ninger}},\ and\ \bibinfo {author} {\bibfnamefont
  {F.}~\bibnamefont {Parmigiani}},\ }\bibfield  {title} {\bibinfo {title}
  {Ultrafast optical spectroscopy of the lowest energy excitations in the
  {M}ott insulator compound yvo3: Evidence for {H}ubbard-type excitons},\
  }\href@noop {} {\bibfield  {journal} {\bibinfo  {journal} {Phys. Rev. B}\
  }\textbf {\bibinfo {volume} {86}},\ \bibinfo {pages} {165135} (\bibinfo
  {year} {2012})}\BibitemShut {NoStop}%
\bibitem [{\citenamefont {Cilento}\ \emph {et~al.}(2018)\citenamefont
  {Cilento}, \citenamefont {Manzoni}, \citenamefont {Sterzi}, \citenamefont
  {Peli}, \citenamefont {Ronchi}, \citenamefont {Crepaldi}, \citenamefont
  {Boschini}, \citenamefont {Cacho}, \citenamefont {Chapman}, \citenamefont
  {Springate}, \citenamefont {Eisaki}, \citenamefont {Greven}, \citenamefont
  {Berciu}, \citenamefont {Kemper}, \citenamefont {Damascelli}, \citenamefont
  {Capone}, \citenamefont {Giannetti},\ and\ \citenamefont
  {Parmigiani}}]{cilento2018}%
  \BibitemOpen
  \bibfield  {author} {\bibinfo {author} {\bibfnamefont {F.}~\bibnamefont
  {Cilento}}, \bibinfo {author} {\bibfnamefont {G.}~\bibnamefont {Manzoni}},
  \bibinfo {author} {\bibfnamefont {A.}~\bibnamefont {Sterzi}}, \bibinfo
  {author} {\bibfnamefont {S.}~\bibnamefont {Peli}}, \bibinfo {author}
  {\bibfnamefont {A.}~\bibnamefont {Ronchi}}, \bibinfo {author} {\bibfnamefont
  {A.}~\bibnamefont {Crepaldi}}, \bibinfo {author} {\bibfnamefont
  {F.}~\bibnamefont {Boschini}}, \bibinfo {author} {\bibfnamefont
  {C.}~\bibnamefont {Cacho}}, \bibinfo {author} {\bibfnamefont
  {R.}~\bibnamefont {Chapman}}, \bibinfo {author} {\bibfnamefont
  {E.}~\bibnamefont {Springate}}, \bibinfo {author} {\bibfnamefont
  {H.}~\bibnamefont {Eisaki}}, \bibinfo {author} {\bibfnamefont
  {M.}~\bibnamefont {Greven}}, \bibinfo {author} {\bibfnamefont
  {M.}~\bibnamefont {Berciu}}, \bibinfo {author} {\bibfnamefont {A.~F.}\
  \bibnamefont {Kemper}}, \bibinfo {author} {\bibfnamefont {A.}~\bibnamefont
  {Damascelli}}, \bibinfo {author} {\bibfnamefont {M.}~\bibnamefont {Capone}},
  \bibinfo {author} {\bibfnamefont {C.}~\bibnamefont {Giannetti}},\ and\
  \bibinfo {author} {\bibfnamefont {F.}~\bibnamefont {Parmigiani}},\ }\bibfield
   {title} {\bibinfo {title} {Dynamics of correlation-frozen antinodal
  quasiparticles in superconducting cuprates},\ }\href
  {http://advances.sciencemag.org/content/4/2/eaar1998} {\bibfield  {journal}
  {\bibinfo  {journal} {Science Advances}\ }\textbf {\bibinfo {volume} {4}}
  (\bibinfo {year} {2018})}\BibitemShut {NoStop}%
\bibitem [{\citenamefont {Eckstein}\ and\ \citenamefont
  {Werner}(2021)}]{eckstein2021}%
  \BibitemOpen
  \bibfield  {author} {\bibinfo {author} {\bibfnamefont {M.}~\bibnamefont
  {Eckstein}}\ and\ \bibinfo {author} {\bibfnamefont {P.}~\bibnamefont
  {Werner}},\ }\bibfield  {title} {\bibinfo {title} {Simulation of
  time-dependent resonant inelastic x-ray scattering using nonequilibrium
  dynamical mean-field theory},\ }\href
  {https://doi.org/10.1103/PhysRevB.103.115136} {\bibfield  {journal} {\bibinfo
   {journal} {Phys. Rev. B}\ }\textbf {\bibinfo {volume} {103}},\ \bibinfo
  {pages} {115136} (\bibinfo {year} {2021})}\BibitemShut {NoStop}%
\bibitem [{\citenamefont {Werner}\ \emph {et~al.}(2021)\citenamefont {Werner},
  \citenamefont {Johnston},\ and\ \citenamefont {Eckstein}}]{werner2021}%
  \BibitemOpen
  \bibfield  {author} {\bibinfo {author} {\bibfnamefont {P.}~\bibnamefont
  {Werner}}, \bibinfo {author} {\bibfnamefont {S.}~\bibnamefont {Johnston}},\
  and\ \bibinfo {author} {\bibfnamefont {M.}~\bibnamefont {Eckstein}},\
  }\bibfield  {title} {\bibinfo {title} {Nonequilibrium-dmft based rixs
  investigation of the two-orbital hubbard model},\ }\href@noop {} {\bibfield
  {journal} {\bibinfo  {journal} {Europhysics Letters}\ }\textbf {\bibinfo
  {volume} {133}},\ \bibinfo {pages} {57005} (\bibinfo {year}
  {2021})}\BibitemShut {NoStop}%
\bibitem [{\citenamefont {Werner}\ and\ \citenamefont
  {Eckstein}(2021)}]{werner2021b}%
  \BibitemOpen
  \bibfield  {author} {\bibinfo {author} {\bibfnamefont {P.}~\bibnamefont
  {Werner}}\ and\ \bibinfo {author} {\bibfnamefont {M.}~\bibnamefont
  {Eckstein}},\ }\bibfield  {title} {\bibinfo {title} {Nonequilibrium resonant
  inelastic x-ray scattering study of an electron-phonon model},\ }\href
  {https://doi.org/10.1103/PhysRevB.104.085155} {\bibfield  {journal} {\bibinfo
   {journal} {Phys. Rev. B}\ }\textbf {\bibinfo {volume} {104}},\ \bibinfo
  {pages} {085155} (\bibinfo {year} {2021})}\BibitemShut {NoStop}%
\bibitem [{\citenamefont {Lang}\ and\ \citenamefont {Firsov}(1962)}]{lang1962}%
  \BibitemOpen
  \bibfield  {author} {\bibinfo {author} {\bibfnamefont {I.}~\bibnamefont
  {Lang}}\ and\ \bibinfo {author} {\bibfnamefont {Y.}~\bibnamefont {Firsov}},\
  }\bibfield  {title} {\bibinfo {title} {Kinetic theory of small mobility
  semiconductors},\ }\href@noop {} {\bibfield  {journal} {\bibinfo  {journal}
  {Sov. Phys. JETP}\ }\textbf {\bibinfo {volume} {16}},\ \bibinfo {pages}
  {1301} (\bibinfo {year} {1962})}\BibitemShut {NoStop}%
\end{thebibliography}
%


\end{document}